\documentclass[12pt,a4paper]{article}
\usepackage{graphicx}
\usepackage{subfigure}
\usepackage[latin1]{inputenc}
\usepackage{amsmath}
\usepackage{amsfonts}
\usepackage{amssymb}
\numberwithin{equation}{subsection}
\def \ud#1{{\underline{#1}}}
\newcommand{\be}{\begin{equation}}
\newcommand{\ee}{\end{equation}}
\def\f{\frac}

\def\non{\\\nonumber}

\newcommand{\beqa}{\begin{eqnarray}}
\newcommand{\eeqa}{\end{eqnarray}}
\newcommand{\p}{\partial}
\begin{document}

\title{
\hfill\parbox{4cm}{\normalsize IMSC/2009/09/12 \\ 
                             }\\
\vspace{2cm}
Properties of CFTs dual to Charged BTZ black-hole
\author{Debaprasad Maity $^a$, Swarnendu Sarkar $^b$, B. Sathiapalan $^a$ 
\\ R.Shankar $^a$,  and Nilanjan Sircar $^a$.\\
$^a$ \small{{\em The Institute of Mathematical Sciences, Taramani,}} \\ \small{{\em Chennai, India 600113}} \\
$^b$ \small{{\em Department of Physics and Astrophysics, University of Delhi,}} \\ \small{{\em Delhi 110
007, India}}
\\ \small{debu,nilanjan,bala,shankar@imsc.res.in, ssarkar@physics.du.ac.in}} }
      
\maketitle
\vspace{-.7 cm}     
\abstract{ 
We study properties of strongly coupled CFT's with non-zero background electric charge in
1+1 dimensions by studying the dual gravity theory - which is a charged BTZ black hole. Correlators of
operators dual to scalars, gauge fields and fermions are studied at
both $T=0$ and $T\neq 0$. In the $T=0$ case we are also able to compare with analytical
results based on $ AdS_2$ and find reasonable agreement. In particular the correlation between log periodicity and the
presence of finite spectral density of gapless modes is seen. The real part of the conductivity (given by the
current-current correlator)  also vanishes
as $\omega \rightarrow 0$ as expected.  The fermion Green's function shows quasiparticle
peaks with approximately linear dispersion but the detailed structure is neither Fermi liquid nor Luttinger liquid
and bears some similarity to a "Fermi-Luttinger" liquid. This is expected since there is a background charge and
the theory is not Lorentz or scale invariant. A boundary action that produces the observed
non-Luttinger-liquid like behavior ($k$-independent non-analyticity at $\omega=0$) in the Greens
function is discussed.}

\newpage
\section{Introduction}

A variety of ill-understood strongly coupled field theories are expected to describe  phenomena
such as High-$T_c$ superconductivity~\cite{Hartnoll:2008vx}~\cite{Hartnoll:2008kx}~\cite{Hartnoll:2009sz}. A technique that
 seems tailor made for this is the AdS/CFT
correspondence~\cite{Maldacena}~\cite{Witten98}, where a string theory in AdS background is dual to a conformal field theory
 on the boundary. 
If the CFT is strongly coupled, then the curvature of the dual gravity background is small
and the massive string excitations do not play much of a role, and so pure gravity is a good approximation. This is the reason
 this technique is ideal for strongly coupled theories.\footnote{ The same fact also makes it difficult to study standard 
continuum QCD, which is asymptotically free and therefore weakly coupled.} 
Historically the correspondence was developed  in Euclidean signature in~\cite{Maldacena}-
~\cite{Skenderis02} to name a few. Some of the references where  real time formalism of the 
correspondence was studied is ~\cite{Balasubramanian:1998de}-~\cite{Iqbal09}

If one wants to study the finite temperature behavior of the CFT, one can study dual backgrounds that are asymptotically AdS
 but have a finite temperature. This could be just pure AdS with a thermal gas of photons or other massless particles, but also
 includes backgrounds that contain a black hole in the interior. There is typically a critical temperature above which the black
 hole is the favored background~\cite{Witten98Therm}. The critical temperature is in fact zero if the boundary is taken to be
 flat $R^n$ rather than $S^n$. This is what happens when we represent the AdS by a Poincare patch. Equivalently there is a
 scaling limit that gives $R^n$ starting from the $S^n$ boundary of global AdS.  Recently such backgrounds with a charged
 Reisner Nordstrom black hole in 3+1 dimensions have been studied to understand 2+1 dimensional CFT's with charge. In
 particular fermionic fields have been studied and interesting  "non-Fermi liquid behavior"~\cite{Senthil08} has been
 discovered~\cite{Lee08}~\cite{Faulkner09}~\cite{Zaanen09}~\cite{Liu09}. 
The same system was studied with magnetic field turned on in ~\cite{Albash:2009wz}
~\cite{Basu:2009qz}.

In this paper we study analogous situations in 2+1 dimensions\footnote{Very recently a similar situation is
studied in~\cite{Hung:2009qk}. Hydrodynamic aspects of $1+1$ field theories using 
$AdS/CFT$ was studied in~\cite{David:2009np}}.
 There are charged BTZ black holes that are
 asymptotically $AdS_3$~\cite{Teitelboim99}. We extract various boundary correlators for scalar operators, currents 
and also fermionic operators in this theory. One of the advantages of studying this system is that since
they describe various strongly coupled CFT's in 1+1 dimension, one has some idea of what to expect in the boundary theory. At the UV end it is dual to
$AdS_3$ and in the IR it flows to another fixed point CFT.
The presence of the background charge breaks Lorentz Invariance and also scale invariance of the boundary theory.
Thus one expects deviations from the Luttinger liquid behavior due to irrelevant perturbations of various types. These deviations
in general change the linear dispersion to a non linear one and the result bears some similarity to a Fermi liquid. A class of these 
theories have been studied in perturbation theory and have been called "Fermi-Luttinger" liquids \cite{FLL}. Thus our holographic
green's functions could be a strong coupling non perturbative version of the Green's function studied in \cite{FLL}.

We should note that  the $AdS/CFT$ correspondence works in full string theory, so ideally we should consider embeddings of $2+1$ dimensional black
hole in full string theory in $9+1$ dimensions. In our present study we assume the existence of such an embedding. By analogy with the best studied example of AdS/CFT correspondence, where Type IIB string theory on $AdS_5 \times S^5$ corresponds
to $SU(N)$ super Yang Mills, one expects that tree level calculation in the bulk corresponds to
the $N\rightarrow \infty$ planar limit of the boundary theory. In our case there is no identifiable parameter $N$. We can take it generically as a measure of the number of species of particles in the boundary theory. Thus when we refer to Luttinger liquid we are actually referring to a $c=\infty$
conformal field theory with a large (infinite) number of scalar fields. Similarly the gravity approximation (neglecting stringy modes) corresponds to large 't Hooft coupling in Yang-Mills. In our case we do not have an action for the boundary theory and hence no well defined notion of coupling constant. However we can assume that the theory describes some non trivial "strong coupling"
fixed point with consequent large {\em anomalous} dimensions. This can be taken as the operational meaning of "strong coupling". 

In the above approximation (planar, strong coupling) used here the full string theory embedding is not required. However the full embedding would in principle determine the charges and dimensions of the operators of the boundary theory, which for us are free parameters. Also to go beyond this approximation would require a knowledge of the embedding.

We study Green's functions corresponding to scalars, fermions and gauge fields at both zero and non zero temperature.
The gauge field calculation gives the conductivity. This is studied at both zero and non zero temperatures. At zero temperature
the small $\omega $ behavior is universal (as shown in \cite{Faulkner09}) and we see that our numerical results
are consistent with this expectation. The same analysis can also be done for fermions and scalars and again there is consistency
with the analytic results at zero temperature and low frequencies. This gives some confidence in the numerical calculations.
   The fermion Green's function does show quasiparticle peaks at specific momenta and these can be identified as Fermi surfaces.
The dispersion relation is approximately linear. The log periodicity in the Green's function are also observed and is consistent with
the analytical expectations. Qualitatively the curves are also similar to the Fermi-Luttinger liquid curves.

We also attempt to reproduce in the boundary theory the intriguing non analyticity in the $T=0$  fermion Green's function 
at $\omega =0$ for {\em any} $k$. We show that it can be explained
if one assumes that there are modes that have their velocity renormalized to zero, interacting with fermions. These could
be thus some localized modes representing impurities.

This paper is organized as follows. In Section 2 we very briefly describe the charged BTZ background. In Section 3 we describe what to expect
from the boundary physics point of view. The paradigm of the Fermi-Luttinger liquid is very useful and is briefly outlined. The numerical results
for the Fermion Green's function is also presented.
Section 4 contains a discussion of the gauge field equation and conductivities are extracted again for zero and non zero
 temperature. Section 5 contains a discussion of the scalar.  Section 6 contains a discussion
of the possible boundary theory that could give rise to these non-analyticities at $\omega =0$. The Appendices contain some
 background material  in the context of the charged BTZ black-hole, 
 that are useful for  some of these calculations and also some details of the calculation.

\section{The Background Geometry:\  Charged $2+1$ dimensional Black Hole}
In our analysis we will consider background of a charged black hole in $2+1$
dimension.  The Einstein-Maxwell action is given by,
\be~\label{EMAction}
S_{EM}=\frac{1}{16 \pi G}\int d^3 x \sqrt{-g} \left( R+\frac{2}{l^2} -4 \pi G F_{\mu \nu} 
F^{\mu \nu} \right)
\ee
Where $G$ is $3D$ Newton constant, $-\frac{1}{l^2}$ is the cosmological constant ($l$ is the
AdS-length). One of the solutions to the above
action is given by the following metric and vector potential after some rescaling(~\cite{Teitelboim99},~\cite{Cadoni08}) (See Appendix (\ref{AppendixMetric}))
\begin{eqnarray}~\label{metric}
ds^2 &=& \frac{1}{z^2} \left[-f(z) dt^2 + \frac{dz^2}{f(z)}+dx^2\right] \nonumber \\
f(z) &= & 1-z^2+\frac{Q^2}{2}z^2 \  ln(z) \nonumber \\
A &=& Q \ ln(z) \ dt
\end{eqnarray}
As $z \to 0$, the metric asymptotes to $AdS_3$ metric and is called boundary. The metric is also singular at $z=1$ called horizon.
The black hole temperature is given by $T=-\frac{f^{\prime}(1)}{4 \pi}=\frac{(1-\frac{Q^2}{4})}{2 \pi}$ ( $|Q|\le 2$ ). 
   
\section{Green's function: Fermion}
A bulk Dirac spinor field $\Psi$ with charge $q$ is mapped to a fermionic operator 
$\mathcal{O}$ in CFT of the same charge and conformal dimension $\Delta=1 \pm m$ . In $2+1$ dimension $\mathcal{O}$ is a chiral spinor. 
By studying the Dirac equation in $2+1$ dimension asymptotically AdS space, we can find 
the retarded Green's Function of the $1+1$ dimensional boundary CFT.~\cite{Faulkner09}
~\cite{Zaanen09}~\cite{Iqbal09}~\cite{Liu09}. We will study behavior for simple  case where bulk fermion is massless ({\it i.e.} $\Delta=1$). Following the calculation 
of ~\cite{Iqbal09}, the boundary Green's function is given by
\be
G_R=\lim_{z \to 0} i \frac{\psi_-(z)}{\psi_+(z)}=\lim_{z \to 0} G(z)
\ee 
where, $\Psi_{\pm}= (1 \pm \Gamma^{\ud z}) \Psi$ and $\Psi_{\pm}=e^{-i \omega t+ i k x} \ \psi_{\pm}(z)$. $G(z)$ defined in the above equation follows
a first order non-linear differential equation which follows directly from bulk Dirac equation,
\begin{eqnarray}\label{Fgreenseq}
z f(z) \partial_z G(z) + G(z)^2 z ( \omega + \mu \ ln(z)-k \sqrt{f(z)})\nonumber \\
+ z ( \omega + \mu \ ln(z)+k \sqrt{f(z)})=0,
\end{eqnarray}
where $\mu=qQ$. In order get the retarded correlation function for the dual fermionic operator
${\cal O}$, we need to impose ingoing boundary condition for $\psi$ 
at the horizon which is equivalent to $G(1)=i$. The boundary condition has to be modified for $T=0$ and $\omega=0$ and is given by
\be
G(z=1)= -\frac{ \sqrt { k^2- \frac{\mu^2}{2} -i \epsilon}}{ (k+\frac{\mu}{\sqrt 2})}
\ee
A detailed analysis for Fermions is given in Appendix (\ref{AppendixGreF}).

\subsection{What to expect}
\subsubsection{Symmetry properties}\label{F_symmetry}
As a consistency check of our numerics we can use the following symmetry properties of the Green's function obtained by direct inspection of the equation of motion (\ref{psifo}) with $m=0$,
$G(\omega,-k,Q,\mu)=-\frac{1}{G(\omega,k,Q,\mu)}$, $G(\omega,0,Q,\mu)=i$, $G(-\omega,-k,Q,-\mu)=-G^*(\omega,k,Q,\mu)$, $G(\omega,k,-Q,\mu)=G(\omega,k,Q,\mu)$. 
\subsubsection{UV behavior}\label{asymptotic}
   Since our background geometry (\ref{metric}) asymptotes to $AdS_3$, in the ultra violet ($\omega >> (T,\mu)$) we expect the effects of finite density and temperature become negligible and it will recover conformal invariance.
If we choose our background geometry as pure $Ads_3$, the Green's function (massless bulk fermion) can be easily obtained as (\cite{Iqbal09}), 
\begin{equation} \label{F_GreensAdS}
G_{AdS}(\omega, k) = I \sqrt{\frac{(\omega + k + i \epsilon)}{(\omega - k + i \epsilon)}}
\end{equation}
where $\epsilon \to 0$. 
This is Green's function for a dimension $1$ chiral operator in $1+1$ dimensional CFT. $Im(G_{AdS})$ or the spectral function  has a symmetry under $(\omega,k) \to (-\omega,-k)$ (``Particle-hole symmetry') and has an edge-singularity along $\omega=k$.
$Im(G_{AdS})$ is zero in the range $\omega=(-k,k)$. Also $\omega \to \pm \infty$, $Im(G_{AdS}) \to 1$ and $G_{AdS}(\omega,k=0)=i$. 

In the ultra violet we expect same scaling behavior ($\Delta=1$) and linear dispersion with velocity unity. Note that the scaling dimension
of the fermionic operator in the boundary is $1$ compared to usual dimension $\frac{1}{2}$ fermionic operators ({\it viz.} electron operator) in $1+1$ dimension. The scaling dimension of the operator in the IR of the boundary theory may
be very different from $1$ as the boundary theory may flow to a different fixed point in IR as described in next section.

\subsubsection{IR behavior}\label{Fermi0ana}
As shown in (\cite{Faulkner09},\cite{Faulkner10}) theories  dual to charged extremal black holes (in $d+1$ dimensions, $d>2$) have a universal IR behavior controlled by the $AdS_2$ region in the bulk. A similar
analysis goes through in $d=2$ case, a brief sketch is given in Appendix (\ref{AppendixAdS2}). At zero temperature, the the background geometry (Appendix (\ref{AppendixAdS2})) is described by $AdS_2 \times \mathbb{R}$ in 
the near horizon limit. From the boundary field theory point of view, although even at $T=0$ the conformal invariance was broken by $\mu$, the theory will have an scale invariance in the IR limit ($\omega<<\mu$) and will
be controlled by an IR CFT dual to $AdS_2$. The IR behavior suggests that the spectral function $A(k,\omega)=Im(G_R) \sim \omega^{\nu_k} $ where  $\nu_k = \frac{1}{2} \sqrt{2(m^2 + k^2) - \mu^2}$. For $\nu_k$ real, the spectral
function vanishes as $\omega \to 0$ and is ``log-periodic'' in $\omega$  if $\nu_k$ is imaginary ($G(\omega e^{n \xi(k,\mu)},k)$ where $n \in \mathbb{Z}$ with $\xi(k,\mu)= \frac{2 \pi}{\sqrt{\mu^2- 2 (m^2+k^2)}}$). There may also
exist particular values of $k=k_F$ where we can have a pole as $\omega \to 0$. The spectral function at those values of $k$ is given by,
\be \label{spectralkf}
A(\omega,k)=\frac{h_1 \Sigma_2}{(k-k_F-\frac{\omega}{v_F}-\Sigma_1)^2+\Sigma_2^2}
\ee  
where $\Sigma_{1,2}$ are real and imaginary part of $\Sigma \sim \omega^{\nu_k}$ respectively.
As mentioned  in (\cite{Faulkner09},\cite{Faulkner10}) that the form of IR Green's function suggests that the IR CFT is a chiral sector of $1+1$ dimensional CFT. 

\subsubsection{Condensed Matter Systems}\label{luttinger}
\begin{figure}[h]
\begin{center}
\subfigure{ \label{F_LL} \includegraphics[scale=.4]{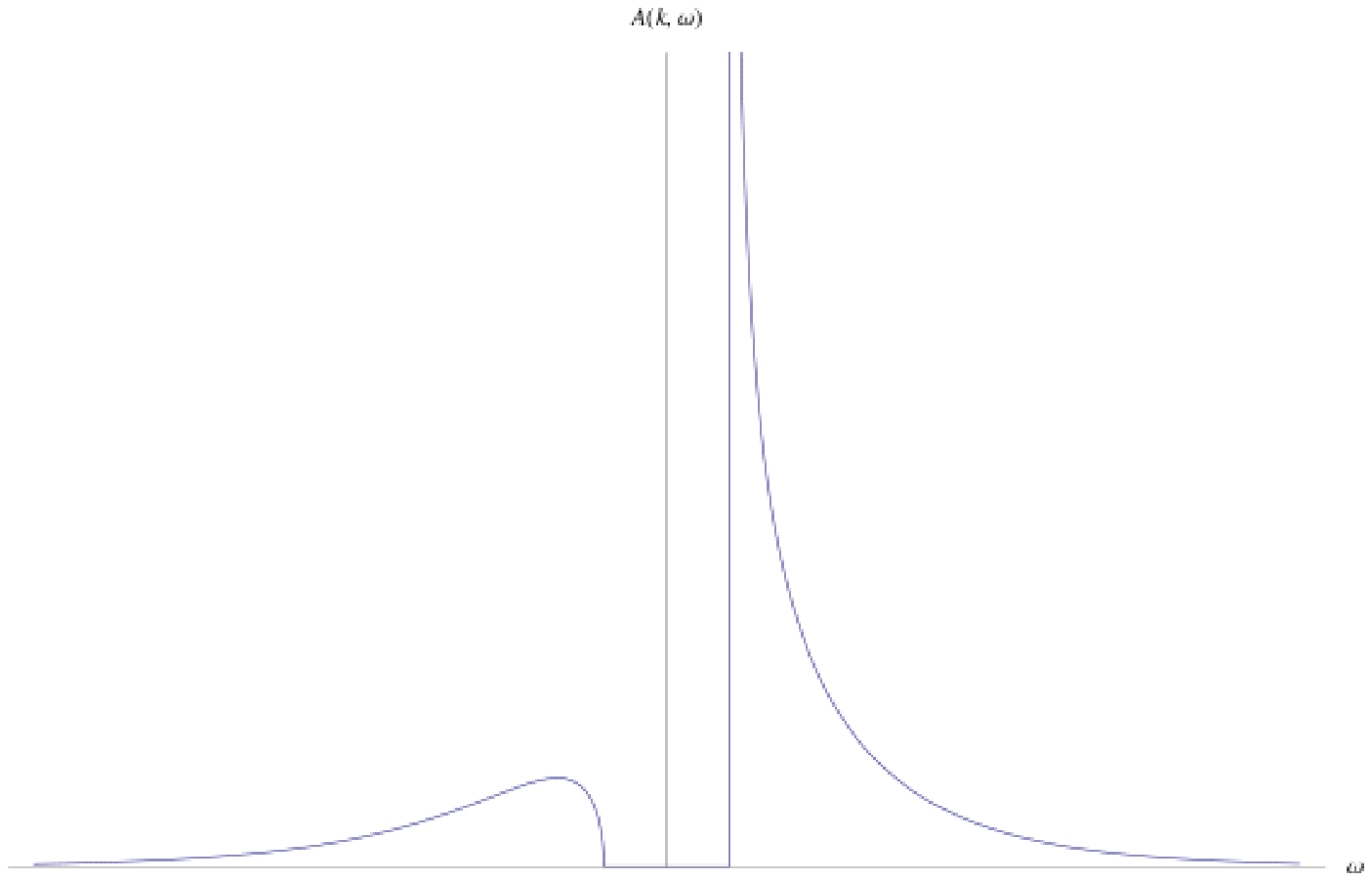}}
\subfigure{ \label{F_LLmod} \includegraphics[scale=.6]{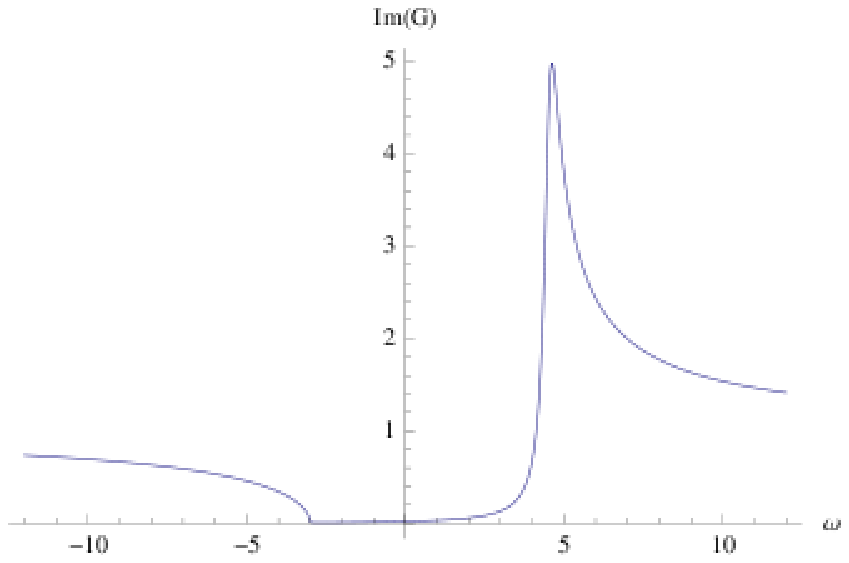}}
\end{center}
\caption{Left:Spectral function of spinless Luttinger Liquid $A(k,\omega)$ vs $\omega$ for $\tilde{k}>0$ and $\gamma=1/2$. It vanishes the range $\omega=(-v_F \tilde{k},v_F \tilde{k})$ 
Right:Spectral function given by eqn.(\ref{F_SpectralLLmodified}) for $k=5$, $\alpha=\beta=1/2$, $\Delta_1=.01,\Delta_2=.2$, $v1=-.6,v2=.9$  }
\end{figure}

For $1+1$ dimension boundary theory one typically expects to get Luttinger Liquid behavior. ``Particle-hole symmetry'($A(\omega,k_F+\tilde{k})=A(-\omega,k_F-\tilde{k})$, $\tilde{k}=k-k_F$ is the deviation of momentum from the Fermi momentum $k_F$ ), 
absence of quasi-particle peak ({\it i.e.} Lorentzian peak known as quasi-particle peak in usual 
Landau Fermi liquid theory is replaced by power law edge singularity) , linear dispersion, spin-charge separation  are the hallmarks of Luttinger liquid in  $1+1$ dimension. In spinless luttinger liquid the spectral function for electron (dimension $\frac{1}{2}$ operator) has a behavior(\cite{LL1},\cite{LL2}) (for $\tilde{k}>0$),
\be~\label{F_SprectralLL}
A(k_F+\tilde{k}) \sim (\omega - v_F \tilde{k})^{\gamma-1} (\omega + v_F \tilde{k})^{\gamma} e^{-a |\omega|}
\ee
$\gamma$ is Luttinger liquid exponent, gives the IR scaling dimension which is related to ``anomalous scaling dimension'' of the fermion operator (See Fig.(\ref{F_LL})).  Deviation from these behavior can be seen in modified Luttinger models {\it e.g.} 
Fermi-Luttinger liquid (\cite{FLL}). In Fermi-Luttinger liquid the dispersion is modified by a non-linear term, and as a consequence the edge singularity for particle  (Fig.(\ref{F_LL})) is replaced
by a Lorentzian peak like Fermi-liquid, but corresponding behavior for hole remains same ({\it ``Particle-hole asymmetry''}). 
The scaling exponent also becomes function of $k$. Very close to the singularity the behavior of the particle spectral function become similar to
(\ref{spectralkf}), which resembles a Fermi-liquid.
We should  note that in our model, 
{\emph both scale invariance and Lorentz invariance is broken for the boundary theory}, so we can expect deviations from Luttinger Liquid. In a Luttinger liquid one could construct correlators like,
\begin{equation}~\label{F_SpectralLLmodified}
G(\omega, k) = \frac{(\omega -v_1  k +i\Delta _1)^\alpha}{(\omega -v_2  k+i \Delta _2)^\beta}~~~~\alpha, \beta>0
\end{equation}
where $\omega,k$ are defined as deviations from Fermi point.
At a fixed point one expects $\Delta _{1,2}=\epsilon \to 0$, and has a edge singularity like (Fig.(\ref{F_LL})). But at a generic point on the RG between
fixed points one expects some finite imaginary part which smooths out the singularity to a peak ({\it viz.} Fermi-Luttinger liquid \cite{FLL}). 
The Green's function have branch cut singularity is along $\omega = v_1 k$ and $\omega = v_2 k$; the peak is along the second line of singularity. The Green's function along the peak becomes,
\be~\label{F_SpectralLLmodified2}
G(\omega )=\frac{(\omega ^2(1-\frac{v_1}{v_2})^2 + \Delta _1 ^2)^{\frac{\alpha}{2}}}{\Delta _2^\beta} e^{i(\alpha \theta - \beta \frac{\pi}{2})}
\ee
where $\tan \theta = \frac{\Delta _1}{\omega (1- \frac{v_1}{v_2})}$. $Im ~G$  clearly increases with $\omega$ ($\alpha, \beta<1$) as long as $\Delta$ does not have an $\omega$
dependence - as at a fixed point Luttinger liquid. In general if $\Delta _1$ is sub linear then one expects $Im ~G \approx \frac{\Delta_1 ^\alpha}{\Delta _2^\beta}$. If $\Delta _1 \approx \omega ^x$ with $x>1$ then
$Im ~G \approx \frac{\omega ^\alpha}{\Delta _2 ^\beta}$.

On the other hand for a Fermi liquid one expects a behavior like (\ref{spectralkf}) but with  $\Sigma \sim \omega^2$. But the IR $AdS_2$ dictates that the scaling exponent of $\Sigma$ is generically different from usual Fermi
liquid. In analyzing the $AdS_3$ example we must keep these points in mind. We expect on general grounds that at least for weak coupling it should behave as a Luttinger liquid. More generally it could go into a massive phase.  
But we do not find evidence of a mass gap in the numerics below. Any deviation from a Luttinger liquid therefore is something noteworthy.
 
\subsection{Numerical results for Fermion}

Since charge $q$ of the bulk field fixes the charge of
boundary operator under global $U(1)$, we must keep $q$ fixed to study a particular boundary operator.
For $T=0$ ($Q=\pm 2$), $\mu= \pm 2 q$, we will study calculate the Fermi-momentum and velocity
for fixed $q$. We will also vary $q$ or $\mu$ to study the ``log-periodicity'' discussed in
section (\ref{Fermi0ana}). For finite temperature case, we will fix $q$ and study the Green's function for
various values of temperature (equivalently $Q$). 

\subsubsection{Zero temperature}\label{fermizero}
\begin{figure}[h]
\centerline{
\begin{tabular}{ll}
 \includegraphics[scale=.5]{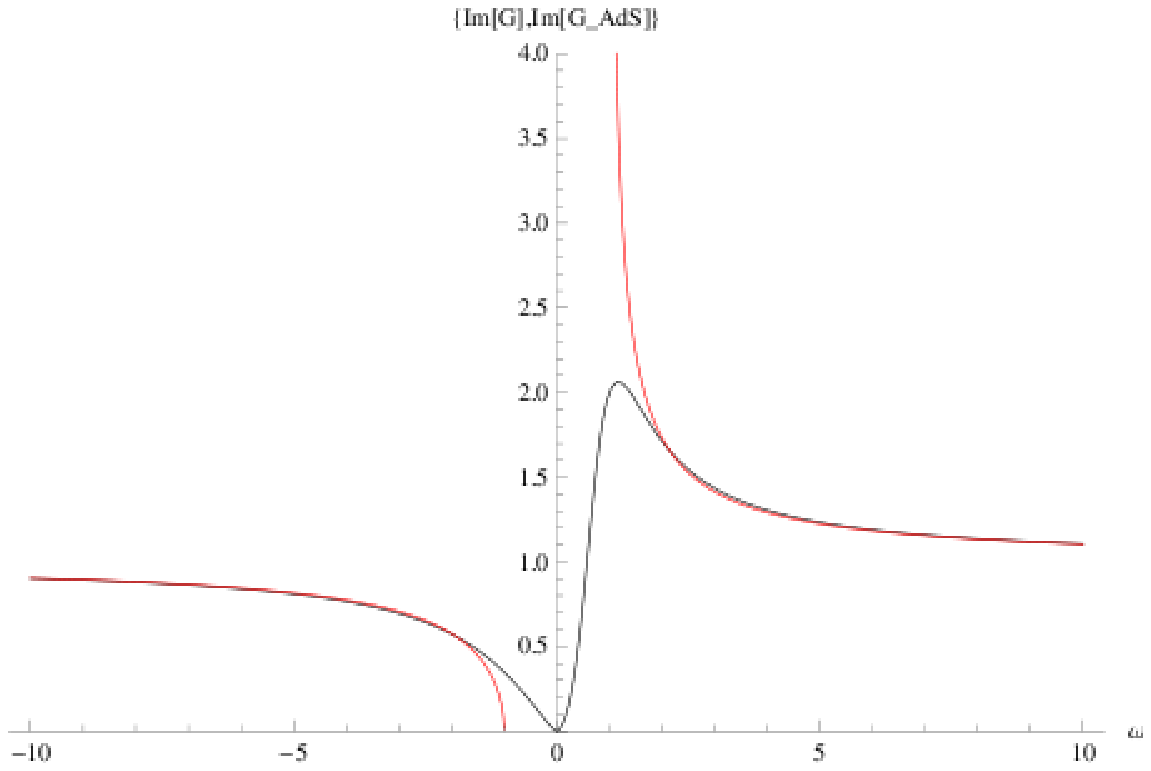} &
\includegraphics[scale=.5]{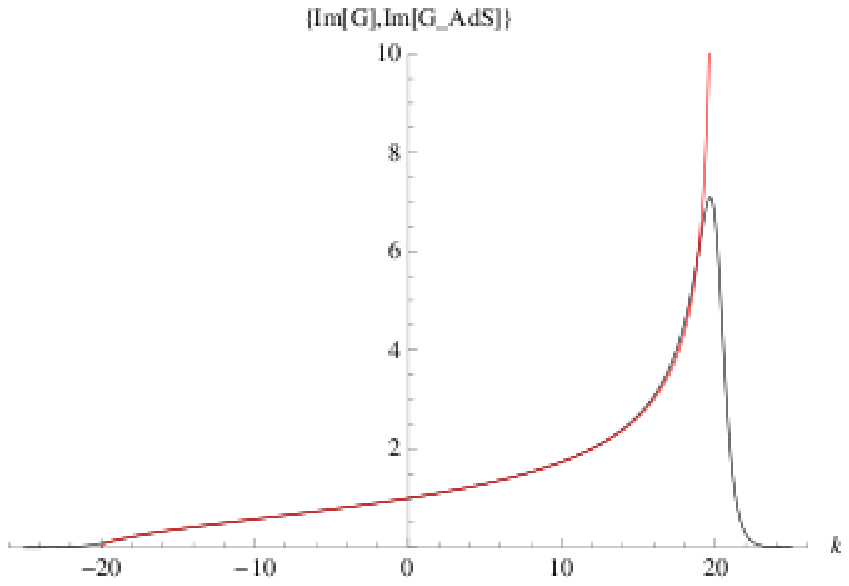}
\end{tabular}
}
\caption{$T=0$: Imaginary part of Fermion Green's function for 
 $q=0$.Left:$k=1$. Right:$\omega=20$}~\label{F_q0}
\end{figure}

\begin{figure}[h]
\centerline{
\begin{tabular}{ll}
\includegraphics[scale=.5]{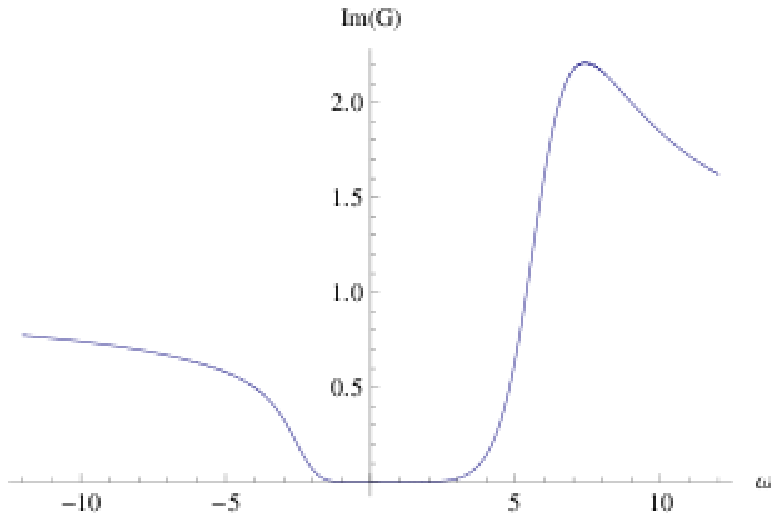} &
\includegraphics[scale=.5]{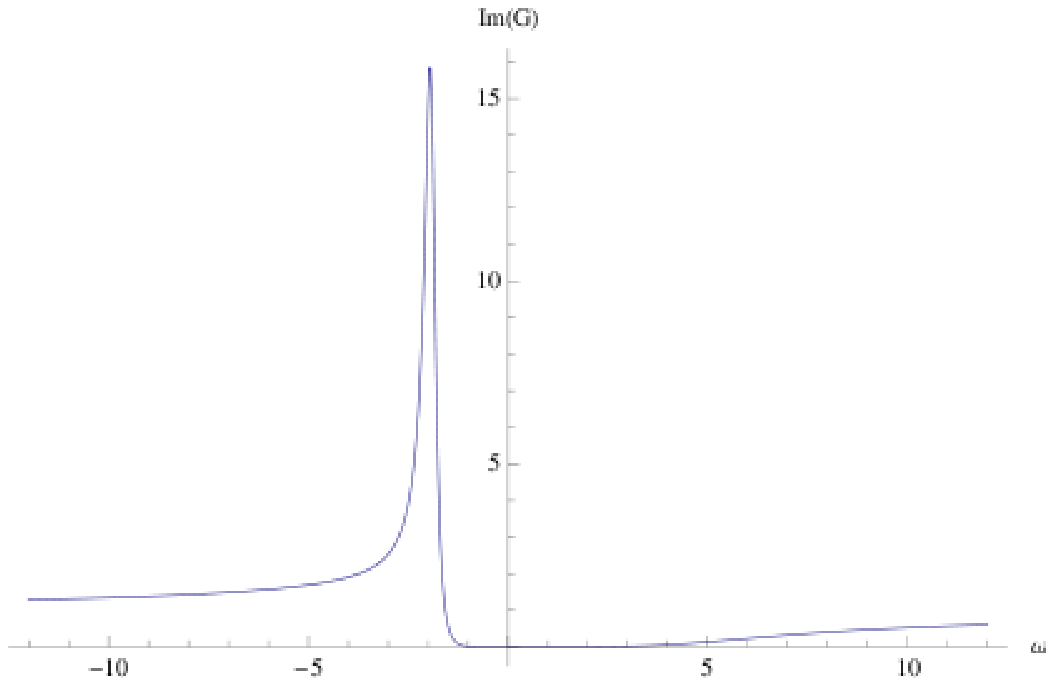}
\end{tabular}
}
\caption{$T=0$: Imaginary part of Fermion Green's function for 
 $q=1/2$.Left:$k=4$ (Compare with what is expected, Figure 1). Right:$k=-4$}~\label{F_qne0}
\end{figure}

\begin{figure}[h]
\centerline{
\begin{tabular}{ll}
 \includegraphics[scale=.5]{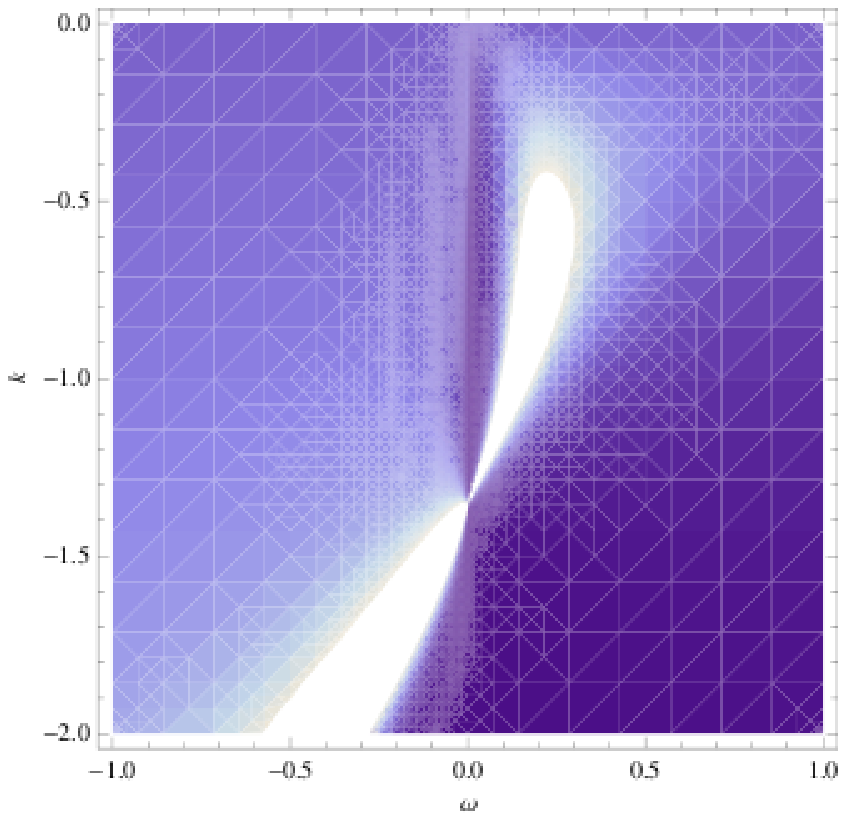} &
\includegraphics[scale=.5]{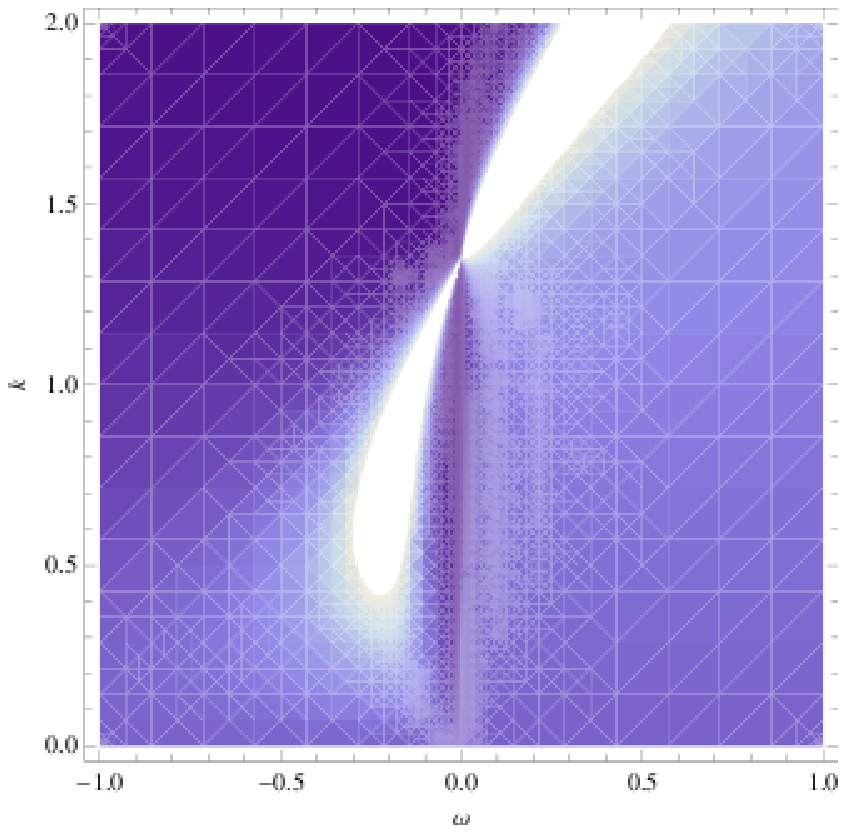}
\end{tabular}
}
\caption{$T=0$: Density plot of imaginary part of Fermion Green's function for 
 $q=1/2$ and $q=-1/2$. Lighter regions have higher value of the function}~\label{F_Density0T}
\end{figure}

\begin{figure}[h]
\begin{center}
 \includegraphics[scale=.7]{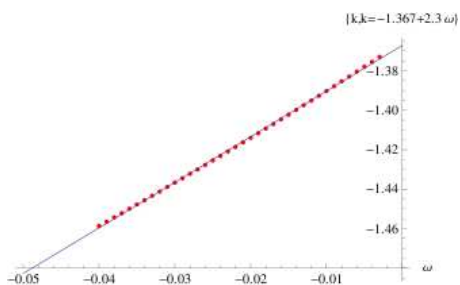}
\caption{$T=0$: Dispersion relation near Fermi point for $q=1/2$ }~\label{F_0Tdisp}
\end{center}
\end{figure}

\begin{figure}[h]
\begin{center}
\subfigure{~\label{F_logp} \includegraphics[scale=.5]{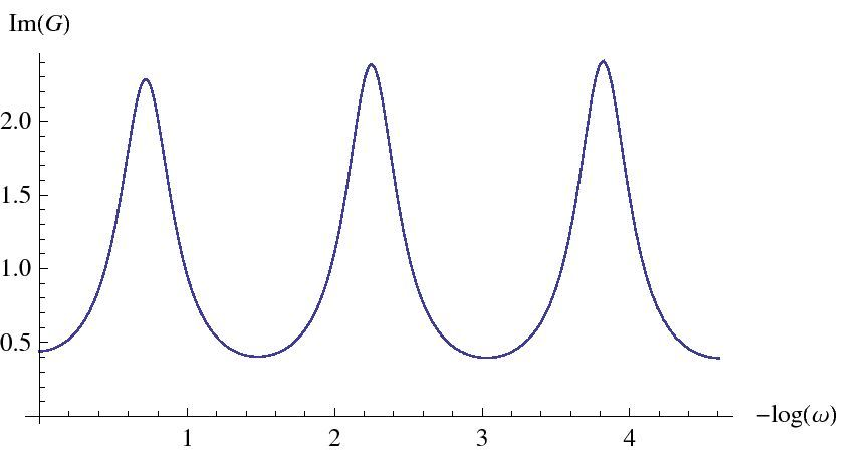}}
\subfigure{~\label{F_linearnear0} \includegraphics[scale=.5]{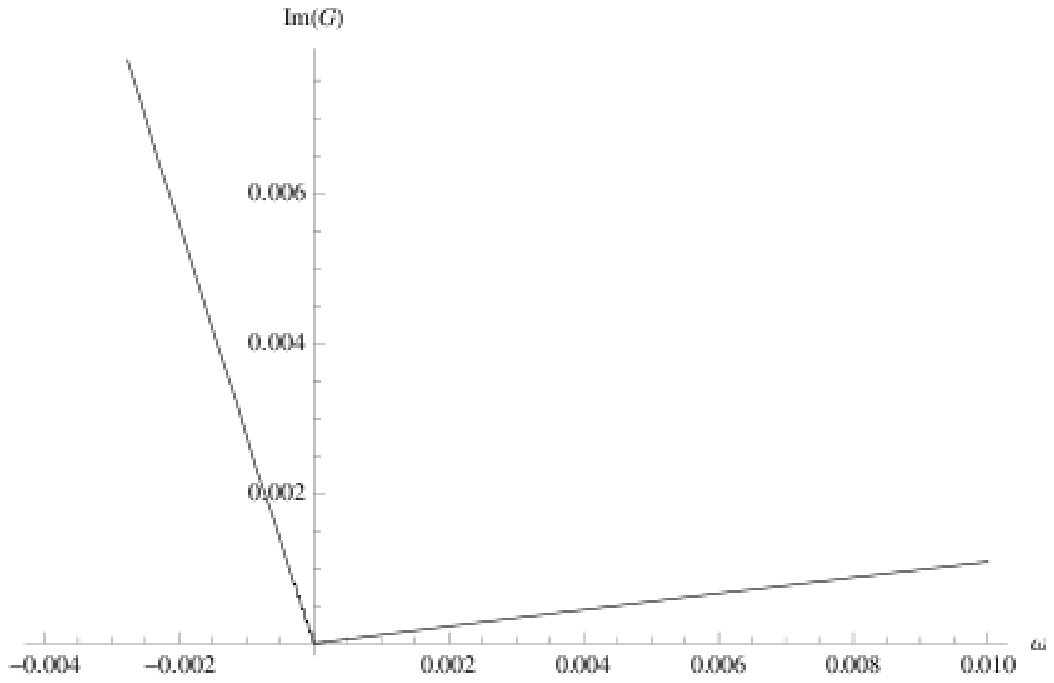}}
\end{center}
\caption{$T=0$: IR behavior of imaginary part of fermion Green's function for 
 Left:$\mu=4$, $k=.5$, Right: $\mu=1$, $k=1$}
\end{figure}

\begin{figure}[h]
\centerline{
\begin{tabular}{ll}
\includegraphics[scale=.5]{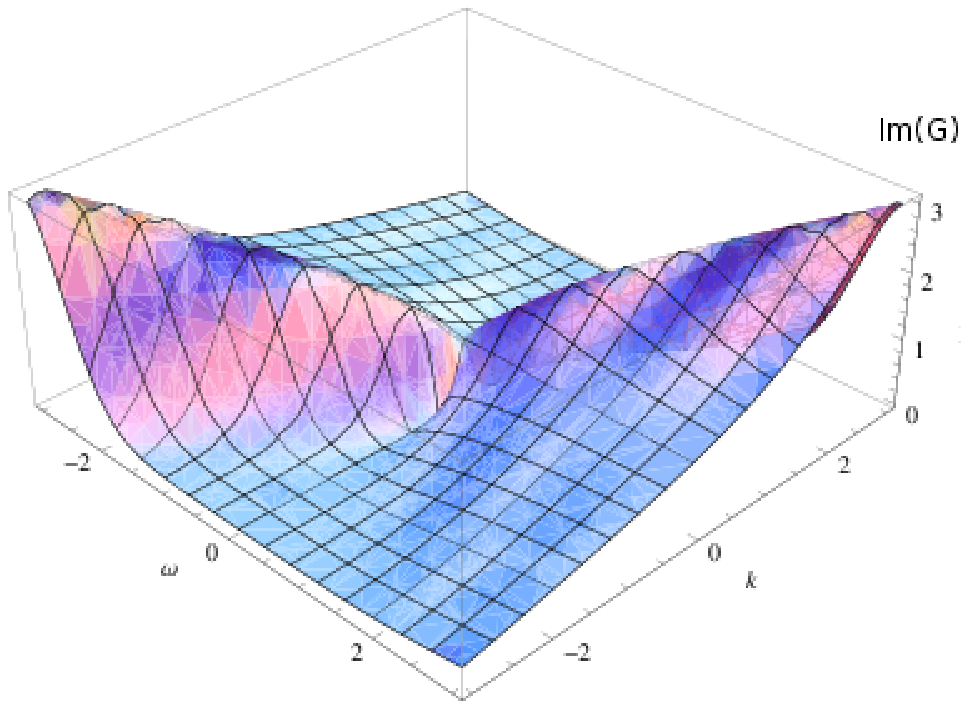}
\includegraphics[scale=.5]{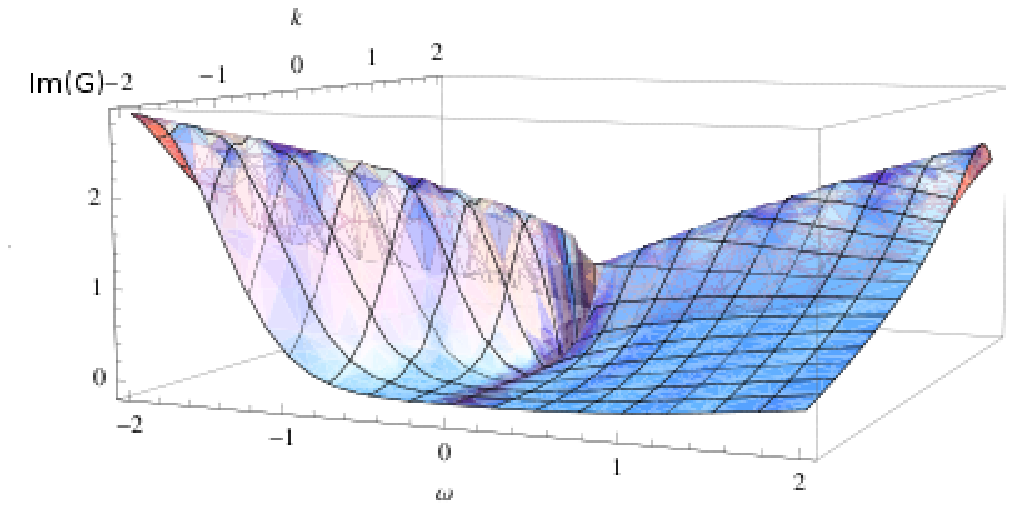}
\end{tabular}
}
\caption{$T=0$: Imaginary part of fermion Green's function for $q=0$ and $q=10^{-1}$ respectively}~\label{F_3dT0q0}
\end{figure}

We will mainly study the qualitative features of zero temperature Green's function as the numerics is not
very precise. Quantitative studies are made at finite temperature where we have very good numerical handle.
\begin{itemize}
\item {\bf UV behavior:}
For $q=0$, the Green's function matches over a large range with pure AdS Green's Function 
(\ref{F_GreensAdS}). For finite $q$, the $G_R \to 1$ as $\omega \to \pm \infty $ for fixed $k$ which implies the UV scaling dimension
of the operator is $1$ (Fig.(\ref{F_q0})).  For fixed $\omega$, the Green's function always matches near $k=0$ which is consistent with symmetry properties (Section (\ref{F_symmetry})). 
The qualitative nature of the peak ($q \ne 0$) matches with Luttinger Liquid Green's function considered in eq.(\ref{F_SpectralLLmodified}) (Compare Fig.(\ref{F_LLmod}) \& (\ref{F_qne0})).
As expected for $k>>\mu$ from pure AdS behavior (eqn.\ref{F_GreensAdS}),  the spectral function should be  zero in the range $\omega+\mu \in (-k,k)$  and a peak near $\omega + \mu \sim k$ (\cite{Liu09}), 
is approximately seen in  Fig.\ref{F_qne0}. 

\item {\bf IR behavior:}
For a fixed $q$ density plot of the spectral function (fig.(\ref{F_Density0T})) shows a sharp {\bf quasi-particle like
peak} at $\omega=0$ and some value of momentum $k=k_F$, called Fermi momentum. At Fermi point the peak height 
goes to infinity and the width goes to zero. The behavior is as described  in section (\ref{Fermi0ana}), this can be seen in modified Luttinger liquids as describe in the paragraph
below the equation (\ref{F_SpectralLLmodified2}).
We expect that the theory flow to a different fixed point in IR. But at  small $q$, along the dispersion curve we see the spectral function has a minima at $\omega=0$  compared to
the quasi-particle like peak, as expected in Luttinger liquid (Fig.(\ref{F_3dT0q0})).
Fermi momentum changes from $k_F$ to $-k_F$ for  $q \to -q$ due to the symmetry (section \ref{F_symmetry}).
For $q=1/2$, $k_F=-1.367$ and $v_F=.43$ obtained from the dispersion plot which is {\bf linear}.(fig.{\ref{F_0Tdisp}}).
If similar analysis was done for $q=1$, where $k_F=1.644$ and $v_F=-.16$. Note $k_F$ changes sign if we change $q$.
Now, as we can see from the expression of scaling dimension of IR operator ${\cal O}_k$ at Fermi momentum ($q=1/2$) turns out to be  $\nu_{k_F} = \frac{1}{2} \sqrt{2 k_F^2-4 q^2}= .83 > \frac{1}{2} $. 
In this regime according to analysis \cite{Faulkner09}  the dispersion relation should be linear. 
Also as described in section (\ref{Fermi0ana}), the spectral function should go to zero if $\nu_k$ real (Fig.(\ref{F_linearnear0}))
and is ``log-periodic'' where $\nu_k$ is imaginary (fig.{\ref{F_logp}}). We found a very good match of the numerical and analytical results for periodicity.
Non-analyticity of Green's function at $\omega=0$ independent of $k$ is not typical of Luttinger liquids and in Section (\ref{boundarytheory}) we have discussed a possible resolution of this.

\item {\bf Particle-Hole (A)symmetry:}  We find ``particle-hole asymmetry'' for $q=\frac{1}{2}$, {\it i.e.} the spectral function behavior is different under reflection at Fermi point (fig.(\ref{F_Density0T})).
This is again very different from Luttinger liquid behavior, but it can be observed in Fermi-Luttinger Liquids as discussed in (\cite{FLL}).
The particle hole symmetry is restored for $\mu=2 q=0$  (fig.(\ref{F_3dT0q0})), and the asymmetry slowly increases with $q$. 

\item {\bf Gapless phase} Also a closer look at density plot (fig.(\ref{F_Density0T})) shows the system is in {\bf gapless} phase which is as expected because in the IR the theory is expected to have conformal invariance.
\end{itemize}

\subsubsection{Finite temperature}
 
\begin{figure}[h]
\centerline{
\begin{tabular}{ll}
 \includegraphics[scale=.7]{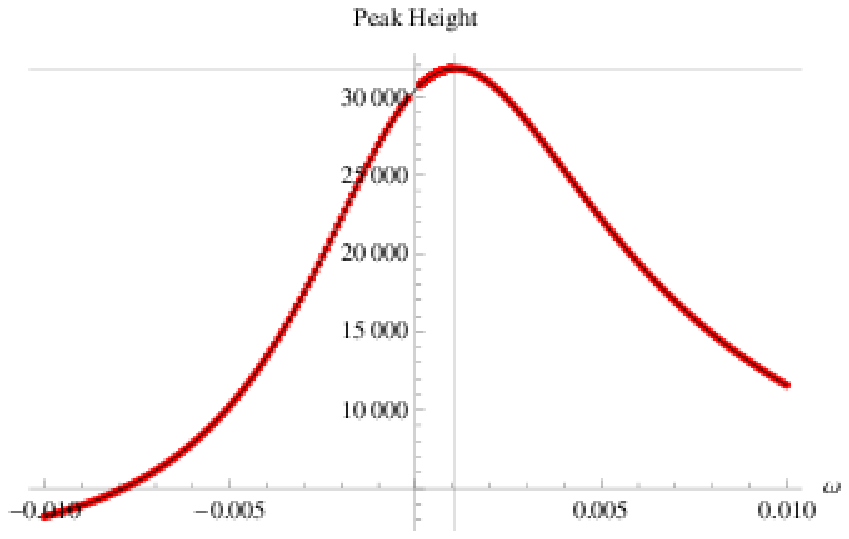} &
\includegraphics[scale=.7]{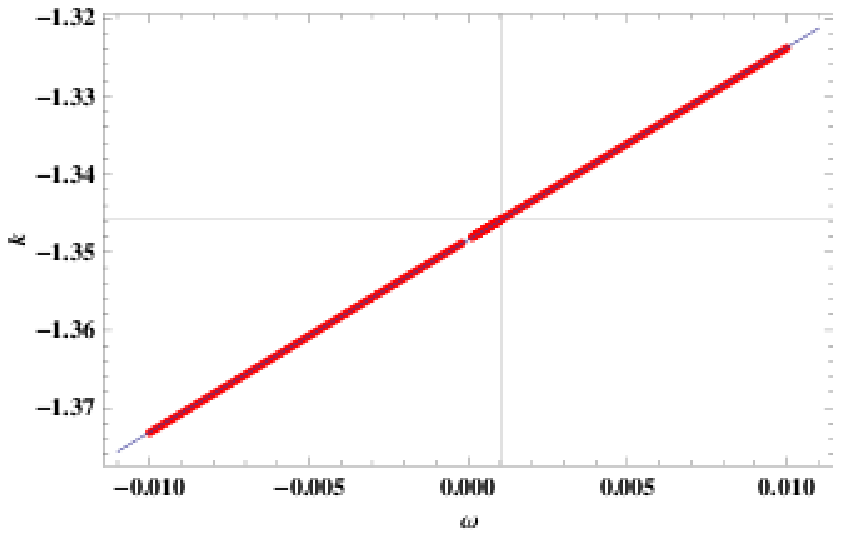}
\end{tabular}
}
\caption{$q=1/2, Q=1.99$ Right: Variation of peak height of imaginary part of (Fermion) Green's function with
frequency shows the Fermi point is shifted to $\omega= 0.00106$. Left: Dispersion is 
linear near Fermi point, $k_F$ is obtained by finding $k$ corresponding to$\omega= 0.00106$ in the linear fit.}
~\label{F_Peakprop}
\end{figure}

\begin{figure}[h]
\begin{center}
\includegraphics[scale=.7]{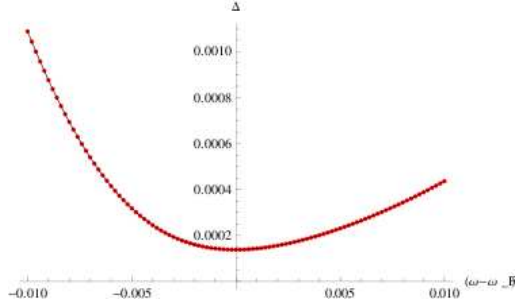} 
\caption{$q=1/2, Q=1.99$. Variation of FWHM with $(\omega-\omega_F)$ (Fermion)}
~\label{F_FWHM}
\end{center}
\end{figure}

\begin{figure}[h]
\centerline{
\begin{tabular}{ll}
\includegraphics[scale=.7]{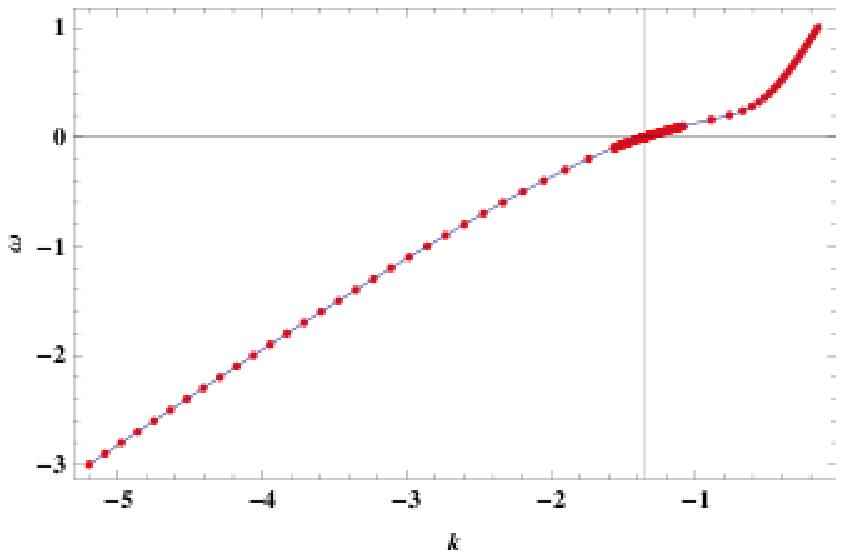}&
\includegraphics[scale=.7]{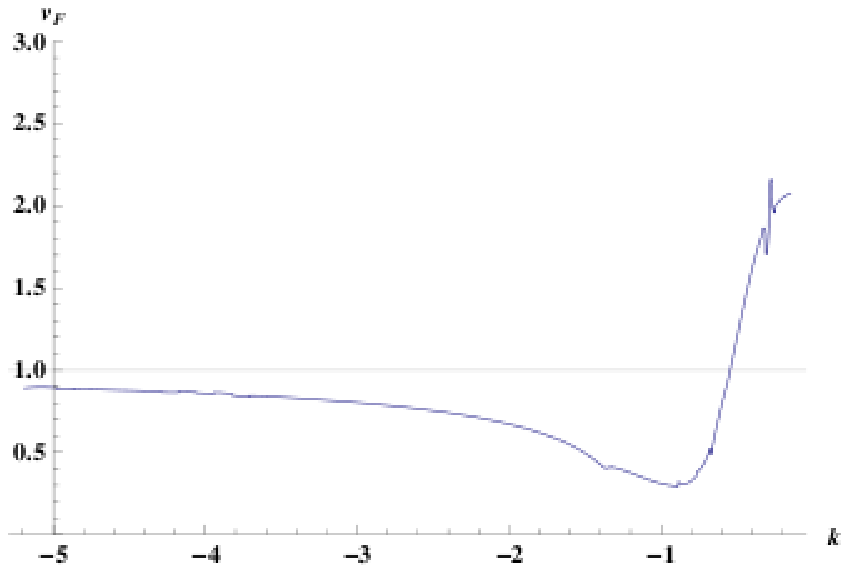}
\end{tabular}
}
\caption{$q=1/2,Q=1.99$. Left: Dispersion curve Right: $v_F$ vs $k$ (Fermion)}
~\label{F_disp}
\end{figure}

The behavior of the spectral function at finite temperature can be summarized as 
follows:
\begin{itemize}
\item {\bf IR behavior:} At sufficiently small temperatures, the quasi-particle like peak still survives. But the width gets broadened with temperature
and also the Fermi frequency shifts to some non-zero value. 
Table (\ref{table1}) shows various peak properties at various temperature with $q=1/2$. Data shows that $\omega_F \ne 0$ is a finite temperature effect and goes
to zero as $T \to 0$. Also $\Delta_F$ (FWHM or ``Full Width at Half Maximum'' at Fermi point) goes to $0$ with $T$. Fermi velocity $v_F$ is independent of temperature.  
The FWHM along the dispersion varies as $\Delta-\Delta_F= A (\omega_F-\omega)^{x}$ where $x=2.25 (\omega < \omega_F),1.8 (\omega > \omega_F)$ at $T=.00159$ (fig.(\ref{F_Peakprop}) ,fig.(\ref{F_FWHM})).
$k_F$ varies linearly with $T$ as $k_F=-1.3671 + 13.49 T$. Which implies at $T=0$,$k_F=-1.3671$ which matches with the zero temperature result.

\item {\bf Dispersion:} The dispersion curve (fig.(\ref{F_disp}),fig.(\ref{F_Peakprop}) )  is linear near the Fermi point, but
deviates from linearity away from the Fermi point. $v_F$ approaches $1$ as $(k-k_F)<<0$ as expected from UV behavior, but $v_F$ becomes large
$(k-k_F)>0$. We expect $v_F$ would go to $1$ for $(k-k_F)>>0$, but we were unable to explore that region as the peak gets very broad in that range and numerical error increases.

\item {\bf Particle-Hole (A)symmetry:} Again in finite temperature case also the spectral function shows ``particle-hole asymmetry'' at low temperature, but the symmetry gets
restored for $Q \to 0$ (large temperature) with fixed $q$. We can conclude that the particle-hole  asymmetry is controlled by the parameter $\mu=qQ$.

\item {\bf Large Temperature:} For large temperature, $Q \to 0$, the Green's function approaches to that for uncharged non-rotating BTZ (UBTZ) as given in (\cite{Iqbal09}) (with $T_L=T_R=\frac{1}{2 \pi}$,
$ G_{UBTZ}=i \frac{\Gamma(\frac{1}{4}-i \frac{\omega-k}{2}) \Gamma(\frac{3}{4}-i \frac{\omega+k}{2}) }{\Gamma(\frac{3}{4}-i \frac{\omega-k}{2}) \Gamma(\frac{1}{4}-i \frac{\omega+k}{2})} $). The quasi-particle peak is completely
lost at large temperature and has a minima along the line of dispersion at $\omega=0$.

\end{itemize}

\begin{table}
\begin{center}
\begin{tabular}{| c | c | c | c | c | c | c |}
\hline
\multicolumn{7}{|c|}{q=1/2} \\
\hline
 Q & T & $\omega_F$ & $k_F$ & $\Delta_F$ & $H_F$ & $v_F$ \\
\hline
1.9 & 0.0155 & -0.00342 & -1.1968 & $4.534 \times 10^{-1}$ & 392 & 0.386\\
\hline
1.95 & 0.007858 & 0.00181 & -1.2712 & $2.735 \times 10^{-3}$ & 1616 & 0.401\\
\hline
1.99 & 0.00159 & 0.00106 & -1.3458 & $1.381 \times 10^{-4}$ & $3.2 \times 10^4$ & 0.406\\
\hline
1.992 & 0.00127 & 0.00089 & -1.3499 & $ 9.34 \times 10^{-5}$ &   $4.7 \times 10^4$ & 0.405 \\
\hline
1.995 & 0.00079 & 0.000597 & -1.3563 & $4.139 \times 10^{-5}$ &  $1.1 \times 10^5$ & 0.404 \\
\hline
1.998 & 0.00032 & 0.000278 & -1.3628 & $8.729 \times 10^{-6}$ &  $5 \times 10^5$ & 0.403 \\
\hline
1.999 & 0.000159 & 0.00013 & -1.3651 & $2.726 \times 10^{-6}$ & $1.6 \times 10^6$ & 0.400 \\
\hline
\end{tabular}
\end{center}
\caption{Peak properties of the Green's function:  \ $\omega_F \to$  Fermi frequency, $k_F \to$
Fermi momentum, $\Delta_F \to $ FWHM at Fermi point, $H_F \to$ Peak Height, $v_F \to$ Fermi velocity}
\label{table1}
\end{table}

\section{Gauge Field: Conductivity}
In this section we consider aspects of the vector field and its perturbation. This gives the current-current correlators of the boundary theory. The vector field is given by the solution
($r$ is the usual radial coordinate and often we will use $z=1/r$.  $r=r_+$ is the location of the horizon).
\be
A_t = -Q ~ln (r/r_+) = Q~ ln (z r_+)  
\ee
If we consider the effective scalar given by $\phi _{eff}=\sqrt g^{tt} A_t = z A_t$ we get
\be~\label{phieff}
\phi_{eff} = Q ~z~ln (z/z_0) + Q~ ln (z_0 r_+) z 
\ee
$z_0$ is an arbitrary normalization scale - denoting violation of conformal invariance.
This is a situation where the eigenvalues are degenerate so $z$ and $z~ln(z)$ are the two independent solution. There is some ambiguity regarding which solution corresponds to the source and which one to the expectation value of the conjugate  operator, 
which in this case is the charge density. Thus if following \cite{Kleb-Witten} we take $Q$ as the value of the source then the charge density is $Q ~ln(z_0 r_+)$.  The analysis of the degenerate case for the scalar does seem to give this - see Appendix A.
On the other hand reversing the roles gives $Q$ as the charge density.

Let us see what thermodynamic arguments give.
Consider the action (\ref{EMAction}) with the metric (\ref{unscmetric}) in $r=1/z$ coordinate,
\be 
ds^2 = -\frac{f(z)}{z^2}dt^2 + \frac{1}{f(z) z^2} dz^2 +\frac{1}{z^2} dx^2
\ee where $f(z)=\frac{1}{l^2}-8 G M z^2+8 \pi G Q^2 z^2 ln(z l)^2$. Also
$F_{tz}=\frac{Q}{z}, ~F^{tz}=-z^3 Q$. The equation of motion also gives:
\be 
R=-\frac{6}{l^2} - 4 \pi G F^2
\ee Thus 
\be 
S= \frac{1}{16 \pi G} \int \frac{d^3x}{z^3} (-\frac{4}{l^2} - 8 \pi G  F^2)
\ee
Thus we get for the charge dependent part of the action :
\be 
S= V_1 \beta Q^2 \int _{\epsilon}^\frac{1}{r_+} ~\frac{dz}{z^3}~z^2 
= -V_1 \beta Q^2 ln( \epsilon r_+)
\ee 
The chemical potential at the boundary is $A_t = \mu = Q ~ln (\epsilon r_+)$. Thus the free energy
is $\beta \Omega =- V_1 \beta  Q^2 ln( \epsilon r_+)= -V_1 \beta \frac{\mu ^2}{ln(\epsilon r_+)}$.
And the expectation value of the charge density is $-\frac{1}{V_1} \frac{\partial \Omega}{\partial \mu}= 2 Q$.
There is no dependence on $r_+$. This suggests that we should treat the coefficient of $z ~ln (z)$ as the
expectation value and not the source. But we will follow the argument given in appendix and consider coefficient of the log term as source.

\begin{figure}
\centerline{
\begin{tabular}{ll} 
 \includegraphics[scale=.65]{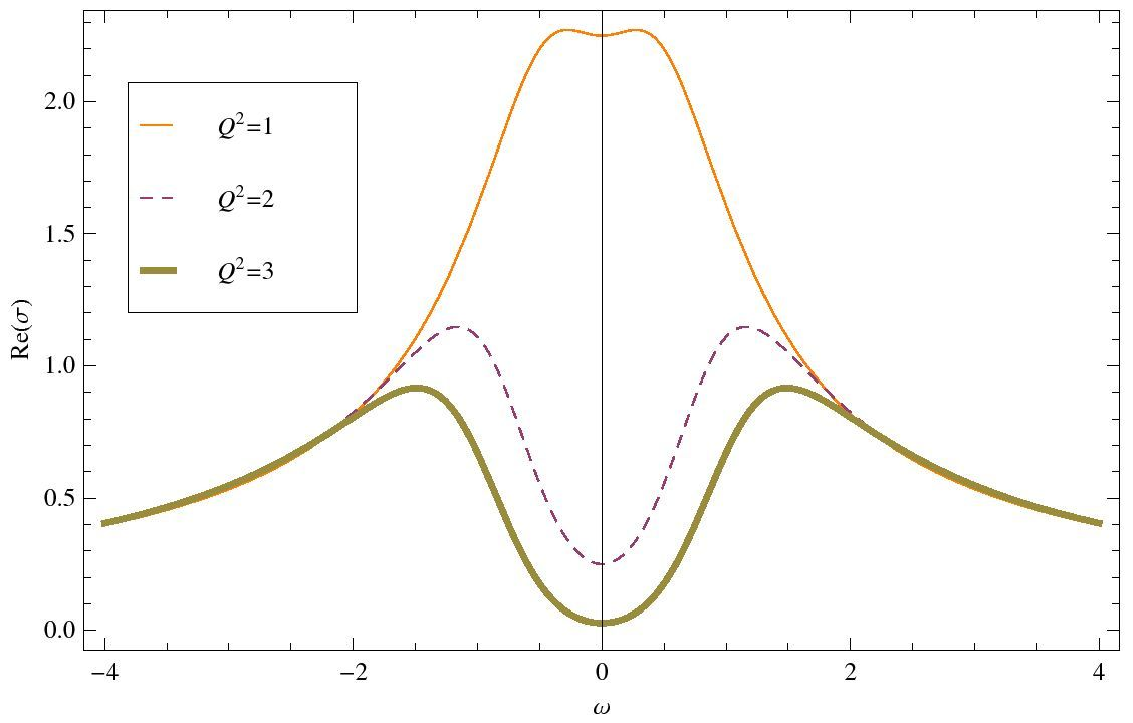} &
\includegraphics[scale=.9]{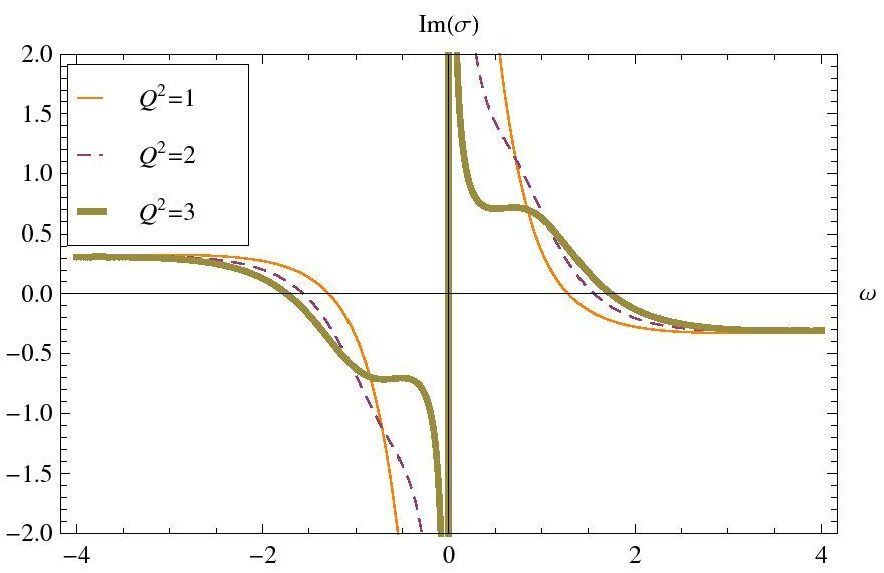}
 \end{tabular}
}
\caption{Conductivity at finite temperature}\label{ConductivityT}
\end{figure}

We now turn to the calculation of the Green's function as a function of the frequency, keeping no $x$ dependence
(i.e.zero wave vector). This is obtained as the ratio of the two solutions in time dependent perturbations $\delta A_x$.
The perturbations in $A$ and $g_{tx}$, which we call $a(z,t)$ and $\delta (z,t)$ respectively obeys  coupled differential equation. We can combine the two equations (Appendix) to get an equation for $a(r,t)$ - which is
(we are assuming a time dependence of $e^{-i\omega t}$ for both $a$ and $\delta$):
\be
f^2 a'' + f ( \frac{f}{z} +f')a' + (\omega^2 - 2Q^2 f)a =0
\ee
Analyzing the leading behavior at $z=0$ we see that the solution has degenerate eigenvalues and is given by $a (z) = a^+ + a^- ~ln (z)$. The Green's function for the conjugate operator $J(x,t)$ is given by the ratio $\frac{a^+}{a^-}$.
( We have to put in-going boundary condition at the horizon, described in detail in the Appendix (\ref{AppendixGreV}))
The ratio between $J^x$ and $\partial _t A_x= -i\omega A_x$ 
is the conductivity $\sigma$. So the Green's function divided 
by $-i\omega$ gives the conductivity (See fig. (\ref{ConductivityT}),(\ref{ConductivityT0})). In the next section we will
discuss about the low energy and zero temperature limit of the 
conductivity. 

\subsection{$\omega \to 0$  limit of Conductivity at $T=0$}
\label{condzero}
\begin{figure}
\centerline{
\begin{tabular}{ll} 
 \includegraphics[scale=.65]{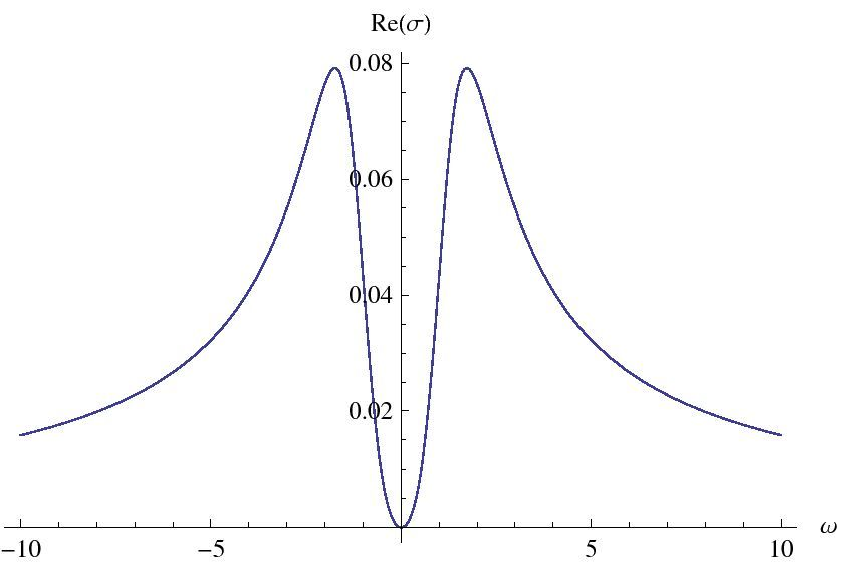} &
\includegraphics[scale=.9]{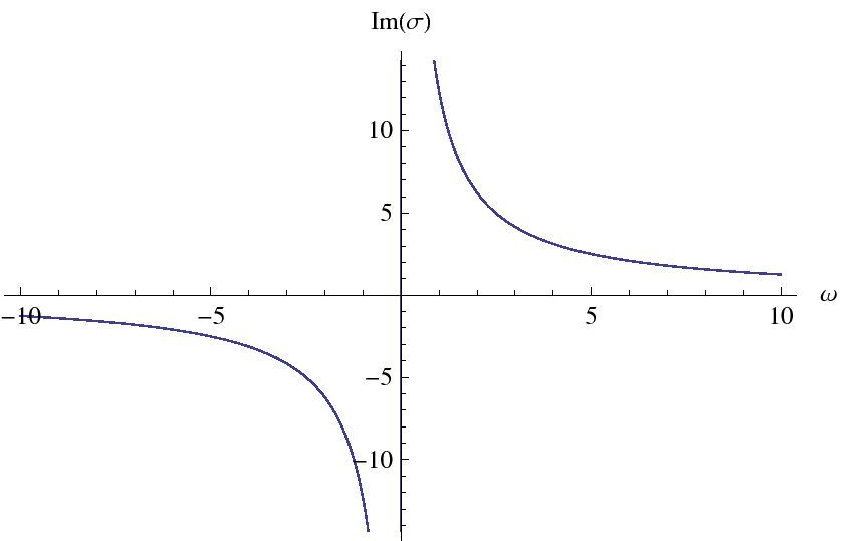}
 \end{tabular}
}
\caption{Conductivity at zero temperature}~\label{ConductivityT0}
\end{figure}

As we have discussed  for the fermion case, for this 
case also we do the same co-ordinate transformation for 
the near horizon limit of the extremal black hole. So, in the 
near horizon $AdS_2$ the equation of motion for the gauge field,
 The leading leading order
$\omega$  behavior of the IR Green's function is given by (see Appendix(\ref{AppendixAdS2})),
\beqa
{\cal G}_R (\omega) = \frac i 3  \omega^3 
\eeqa. 
So, as mentioned earlier the real part of the conductivity
is
\be
Re \sigma (\omega \rightarrow 0) = \lim_{\omega \rightarrow 0} \frac 1 {\omega}
Im {\cal G}_R(\omega)\sim \omega^2
\ee
So, we see  that a dimension 2 IR CFT operator determines the limiting 
behavior of the optical conductivity in low frequency  limit of
boundary $(1+1)$ dimensional field theory. This behavior should be contrasted with behavior of doped Mott insulator (\cite{Giamarchi96}), where the real part of conductivity goes as
$\omega^3$. In addition, the imaginary part of $\sigma$ has a pole at $\omega=0$ which is also clear from the plot (Fig.\ref{ConductivityT0}),
\be
Im \sigma(\omega \rightarrow 0)\propto \frac 1 {\omega}
\ee
Now, from the well known Kramers-Kronig relation 
\be
Re(\sigma) \propto \delta (\omega)\ee at $\omega =0$. From the numerical analysis (fig.\ref{ConductivityT0}) we were not able to see the delta function peak. But this is a natural
expectation that for any translational invariant theory the DC conductivity (for $\omega =0$), should be infinity.  

\section{Complex Scalar}

In this section we consider the complex scalar field in the background geometry (\ref{metric})  and 
derive the Green's function numerically for the dual operators in the boundary. 
The equation of motion of the scalar is given by,
\beqa\label{scalareqn}
A(z)\phi_k^{''}(z)+B(z)\phi_k^{'}(z)+V(z)\phi_k(z)=0
\eeqa

\beqa
A(z)&=&z^2f(z)\non
B(z)&=&z[f^{'}(z)z-f(z)]\non
V(z)&=&z^2[-k^2+\f{1}{f(z)}(\omega +qQln(z))^2]-m^2\nonumber
\eeqa

The solution to the equation (\ref{scalareqn}) near the boundary, $z\rightarrow 0$, for non-zero mass of the scalar field 
behaves as, (we have dropped the log term coming from the potential term, $V(z)$ in (\ref{scalareqn}) assuming a finite mass for the scalar field\footnote{In the following numerical computations we have set $m^2=1/4$.}) 

\beqa
\phi(z)=a^-(k,\omega)z^{d-\Delta}+a^+(k,\omega)z^{\Delta}
\eeqa

Here $d=2$ is the boundary space-time dimension and $\Delta=d/2+\sqrt{(d/2)^2+m^2}$. As per \cite{Kleb-Witten} the boundary
Green's function is given by, $a^+/a^-$. The solution for $\phi(z)$ is obtained numerically and 
the boundary Green's function can be written as,

\beqa\label{gfscalar}
G(k,\omega)&=&\f{a^+(k,\omega)}{a^-(k,\omega)}\non
&=& \f{z^{d-2\Delta}}{2\Delta-d}\left[(\Delta-d)+z \f{\phi^{'}_k(z)}{\phi_k(z)}\right] \mbox{\hspace{0.1in};\hspace{0.1in}} z\rightarrow 0
\eeqa

We solve equation (\ref{scalareqn}) numerically with in-going boundary condition at the horizon, and obtain the Green's function using  (\ref{gfscalar}). The procedure is described briefly in the Appendix (\ref{AppendixGreS}).

\subsection{ $\omega \rightarrow 0$ limit of Scalar Green's Function at Zero Temperature}

Following a calculation similar to Fermion (see Appendix (\ref{AppendixAdS2})), the Green's functions in the $AdS_2$ goes as $\omega ^{2\nu }$ where $\nu = \sqrt{\frac{1}{4}+
(\frac{m^2+k^2}{2}-\frac{q^2 Q^2}{4})}$ . This is just the same as the 
analysis of \cite{Faulkner09}, who use this to further 
conclude that the imaginary part of the full Green's 
function has this behavior. If $\nu$ is not real one has also the interesting periodicity in $\log(\omega)$.

We should note that the effective mass of the charged scalar field 
in $AdS_2$ in the presence of the gauge field is $(\frac{m^2+k^2}{2}-\frac{q^2 Q^2}{4})$.
The condition for log-periodicity implies that this effective mass is lower than the
BF bound in $AdS_2$ which is $-1/4$. This means that for the choice of the parameters
which exhibit log-periodicity of the Green's function the scalar field is unstable.

\begin{figure}\centerline{
\begin{tabular}{ll}
 \includegraphics[scale=.8]{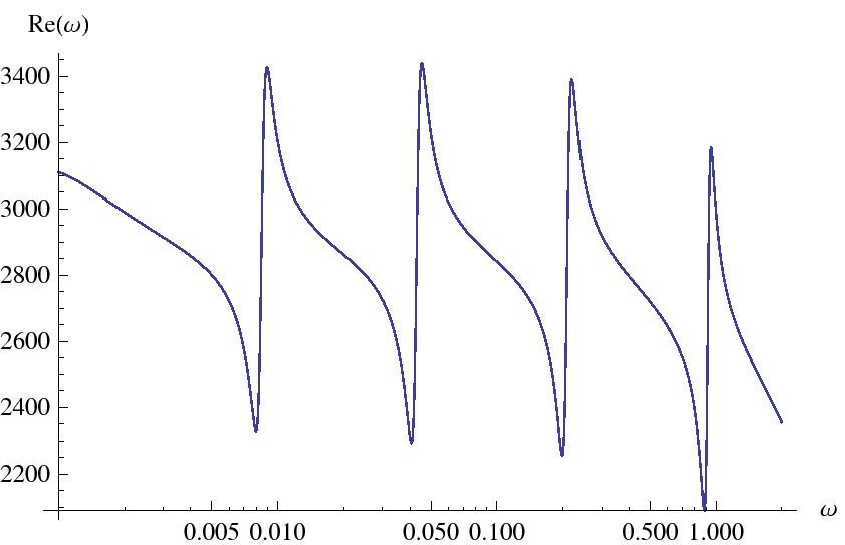} &
\includegraphics[scale=.8]{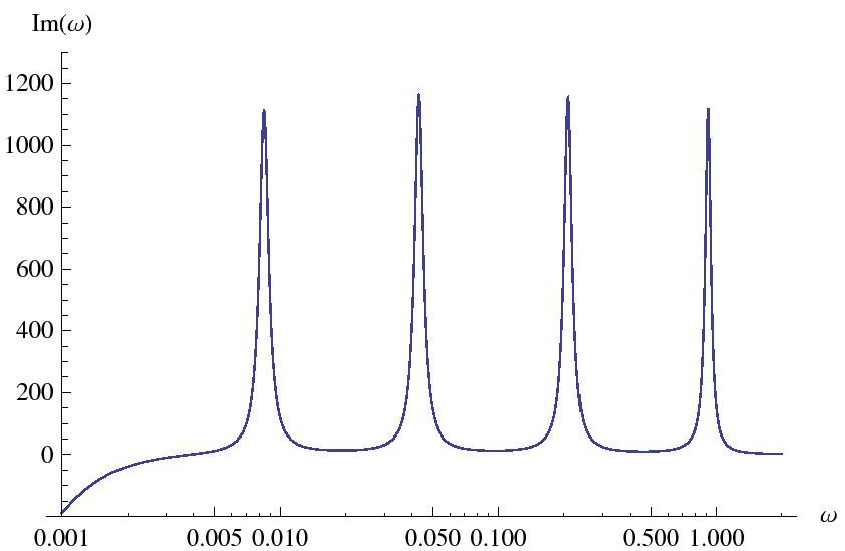}
\end{tabular}
}
\caption{Re$G(\omega)$ and Im$G(\omega)$ for $k=0$, $q=2$ and $m^2=1/4$. The plots are with respect to $\log(\omega)$, however the markings 
on the horizontal axis are that of $\omega$.}
\label{scalarlogp}
\end{figure}

Figure \ref{scalarlogp} shows that the log-periodic behavior (in $\omega$) of the real and the
imaginary parts of the Green's function. We have plotted these
for $q=2$, $k=0$ and for $m^2=1/4$. Note that this value of mass is well above the BF bound for the
scalar field in $AdS_3$ (which is $-1$), although the effective mass in $AdS_2$ is unstable. The scalar
field though asymptotically stable is unstable in the near-horizon $AdS_2$ region. 
We expect that this instability will
lead to the condensation of the scalar field. In the case of charged scalars in $AdS_4$ 
this instability leads to a transition to a hairy black hole phase which in the dual theory
has been identified as the transition to a superconductor phase (see \cite{Horowitz2010} for nice review). For our case we shall analyse 
this aspect in our upcoming work.
   
The imaginary part of the Green's function becomes negative for 
smaller values of $\omega$. We presume that this is due to instabilities occurring from 
the fact that the effective $AdS_2$ mass for the scalar is tachyonic in the region of parameter 
values where the Green's function shows log-periodicity. However the log-period that appears in the 
plot matches with the analytical value which in this case is $1.65$.  

\subsection{Numerical results}
Let us now note some general properties of the Green's functions before discussing the numerical results. 
 From the differential equations it can be seen the Green's function is even in $k$, $G(\omega,k,qQ)= G(\omega,-k,qQ)$. 
For $Q=0$, complex conjugation of the Green's function has the same effect as $\omega \rightarrow -\omega$. This implies
that Im $G(\omega,k)=-$ Im $G(-\omega,k)$, and Re $G(\omega,k)=$ Re $G(-\omega,k)$.  
For non-zero values of $qQ$, this symmetry is lost due to the presence of the 
covariant derivative, however, $G(\omega,k,qQ) = G^*(-\omega,k,-qQ)$. The asymmetric behavior is visible in all the plots for nonzero 
$Q$. We now turn to our numerical results. 

\begin{figure}
\centerline{
\begin{tabular}{ll}
 \includegraphics[scale=.8]{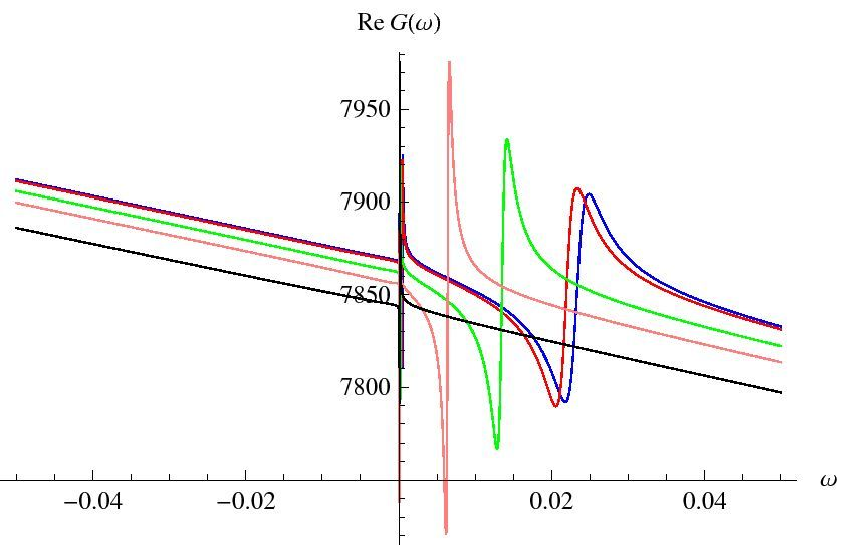} &
\includegraphics[scale=.8]{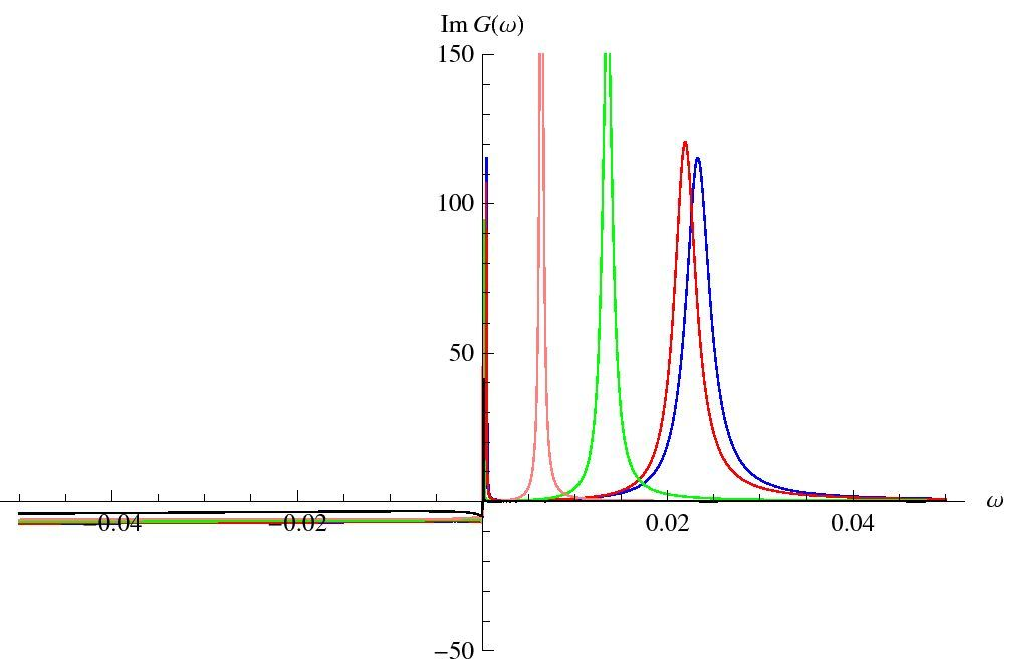}
\end{tabular}
}
\caption{Re$G(\omega)$ and Im$G(\omega)$ for $Q=2$ or $T=0$ and for the following values of $k$: $0.1$ (Blue), $0.2$ (Red), $0.5$ (Green), 
$0.7$ (Pink), $1$ (Black)}
\label{zero_temp}
\end{figure}

\begin{figure}
\centerline{
\begin{tabular}{ll}
 \includegraphics[scale=.8]{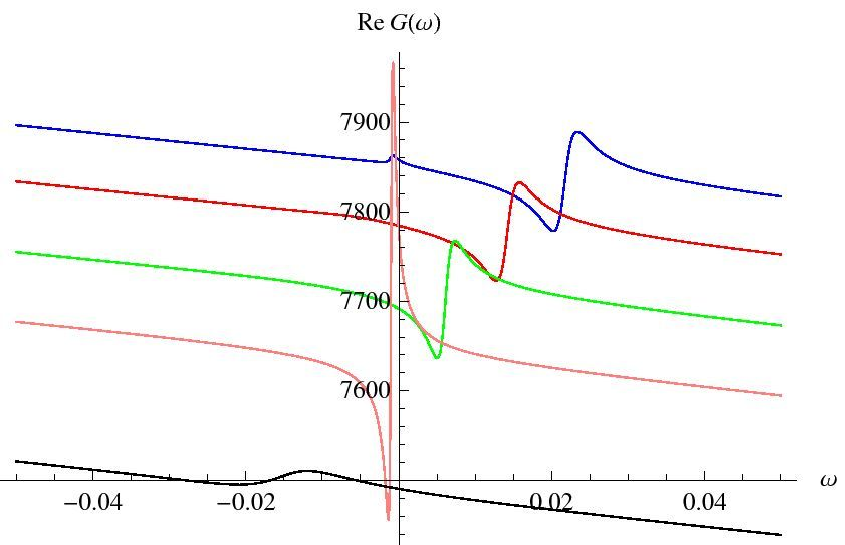} &
\includegraphics[scale=.8]{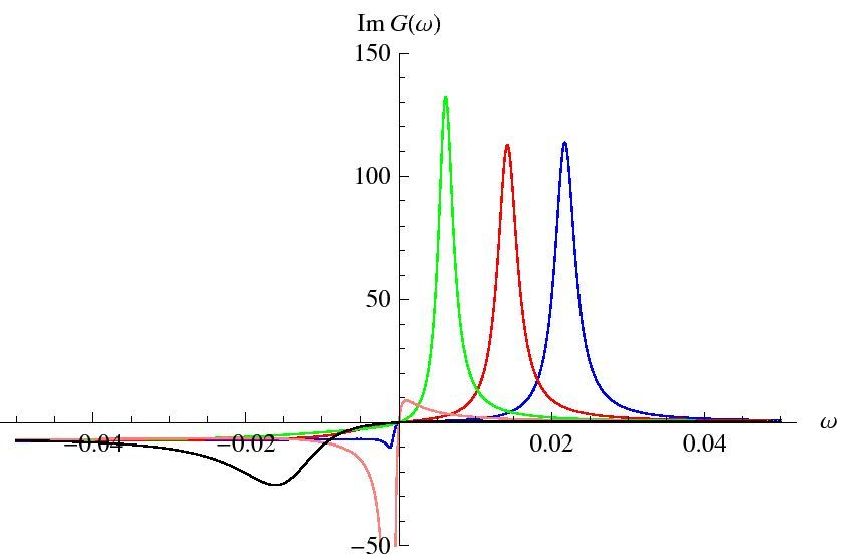}
\end{tabular}
}
\caption{Re$G(\omega)$ and Im$G(\omega)$ for $k=0$ and for the following values of $Q$: $1.998$ (Blue), $1.99$ (Red), $1.98$ (Green), 
$1.97$ (Pink), $1.95$ (Black)}
\label{finite_temp}
\end{figure}

\noindent
{\bf Zero temperature:} Figure \ref{zero_temp} shows the real and imaginary parts of the boundary Green's function for the extremal 
case, $Q=2$ or $T=0$, $q=1$ and for various values of $k$. At zero temperature, the Green's function has peaks lying between $\omega=0$ and $\omega=0.05$.
These peaks shift towards $\omega=0$ as $k$ is increased which  
disappear for larger values of $k$.  The plot (Figure \ref{zero_temp}) shows a range of $k$ for which the peaks exist. We would like to 
also point out that the values of the parameters considered here lie in the region where the Green's function is periodic in $\log(\omega)$ for small $\omega$. This is discussed in more detail in the next section.
\\
\noindent
{\bf Finite temperature:} For a fixed value of $k$, the peaks lying in the positive $\omega$ region smoothen out beyond a 
particular temperature. Figure \ref{finite_temp} shows the plot for the real and the imaginary parts of 
the Green's functions for various values of
$Q$ or $T$ and $k=0$ and for $q=1$. The Green's function is very sensitive to variations in temperature near the zero temperature region. The number of peaks 
reduce to one in the 
finite temperature regions for a particular value of $k$. Eventually this peak smoothens out as the temperature is increased further. At the highest temperature $Q=0$, the curves (not shown in the figure) have the symmetries 
mentioned at the beginning of this section. For large values of $\omega$, the functions are monotonically increasing or
decreasing. Like the zero temperature case the peaks shift towards the origin of $\omega$ and then disappear for larger values of $k$. 
In fact, like at zero temperature there exists a range of $k$ for which there are peaks. 
This window of $k$ for which the peaks exist shrinks as the temperature is increased.  
\\
\noindent
{\bf Large $\omega$:} In the large $\omega$ limit for fixed $k$ the Green's function should behave as, $|G(\omega)|\sim \omega^{\alpha(m)}$,
With $\alpha(m)=2\Delta-d$. This scaling behavior can be verified numerically for our Green's function. We get the following results from our numerical
computation at finite temperature: $\alpha(0.5)=2.229$, $\alpha(1)=2.830$, $\alpha(1.5)=3.627$. These may be compared to the analytical values: $\alpha(0.5)=2.236$, $\alpha(1)=2.828$, $\alpha(1.5)=3.605$. We have verified that results correct to the first decimal place can be obtained for the real and imaginary parts of the Green's function both for finite and zero temperatures. Similarly the same scaling can be checked for large $k$ with fixed 
finite $\omega$. We get the numerical answers correct to the first decimal place as above.

\section{Non Fermi Liquid Behavior in the Boundary Field Theory} \label{boundarytheory}
Since our boundary theory lies on an RG trajectory connecting a UV fixed point CFT dual
to $AdS_3$ and a CFT in the IR (because it is gapless) we can expect it to be described by
the action of a Luttinger liquid like theory modified by the addition of some irrelevant
perturbations. These terms are allowed
because the background charge density (implied by the chemical potential $A_0(z=0)$) breaks
scale invariance. It also breaks Lorentz invariance, which means that one cannot assume that
all propagating modes propagate with the same velocity. The effect of these irrelevant terms have been investigated in \cite{FLL}
in perturbation theory. One expected effect is that the linear dispersion relation is corrected by non linear terms. This has the
consequence that the interaction terms induce a finite imaginary part to the self energy correction and in some approximation
it introduces a Lorentzian peak modifying the edge singularity of the Luttinger liquid. This is called a Fermi-Lutttinger liquid
in \cite{FLL} and was discussed in Sec. 3. 

The holographic analysis also showed an interesting non Luttinger behavior in that the zero frequency behavior is peculiar. It
has a singularity at $\omega =0$ that is $k$-independent. In this section we attempt to reproduce this in the boundary theory.

Let us consider the following action:
\be
S=\int dx dt [i\bar \psi \gamma ^\mu \p _\mu \psi - \frac{1}{2} \p _\mu \phi \p ^\mu \phi]
\ee

$\psi$ is the quasiparticle fermion field whose Green's function we are interested in determining.
It has dimension half unlike the fermion operator that was studied in the last section, which had dimension one.
 Nevertheless this is a matter of detail and will not affect the main point of this section.
 $\phi$ is the bosonized version of some other charged fermion, $\Psi$.
 They are related by $\bar \Psi \gamma^\mu \Psi = \epsilon ^{\mu \nu}\p_\nu \phi$.

The above action assumes Lorentz invariance and we have set $v=1$. But because Lorentz invariance
is absent one can expect the two quasiparticles to have different velocities:
\be
S=\int dx dt [i\bar \psi (\gamma ^0 \p _0 + \gamma^1 v_F \p _1) \psi -
 \frac{1}{2}( \p _0 \phi \p ^0\phi +v_S^2\p _1\phi \p ^1 \phi)]
\ee

Both $v_S$ and $v_F$ can get renormalized by interactions. For instance adding a charge-charge interaction
 term for $\Psi$ adds $(\p_x\phi)^2$ to the action. This modifies $v_S$. 
 In fact we will assume that $v_S$ is renormalized to zero. This will be crucial as we will see below.

We now add an interaction term \footnote{ Instead of $\bar \psi \psi$ one can add $\psi ^\dag \psi$ which would correspond to a density. 
This term is allowed since we do not have Lorentz invariance and also has an interpretation of some spin density interacting with some projection of a magnetic field, for instance. We thank Thomas Vojta for suggesting this.}:
\be
\Delta S = g\int dx dt ~i \bar \psi \psi cos~ \beta \phi
\ee

We can perform a Wick rotation ($i t _M= t_E$) to Euclidean space for calculations.
Using the fact that $\langle \phi (x,t) \phi(0,0)\rangle =-\frac{1}{4\pi}ln~(r^2+a^2)$ we see
that $cos ~\beta \phi = (a)^{\frac{\beta^2}{4\pi}}:cos \beta \phi:$ and has dimension $\frac{\beta^2}{4\pi}$.
The interaction term that has been added is thus irrelevant when $\frac{\beta^2}{4\pi}>1$.
One also sees that
\be
\langle :cos \beta \phi (x,t)::cos \beta \phi (0,0):\rangle =\frac{1}{2} \frac{1}{(x^2+t^2)^{\frac{\beta^2}{4\pi}}}
\ee
The propagator in momentum space can be obtained by dimensional analysis and is (in Minkowski space)
$G(\omega,k)_{cos\beta \phi} \approx [\omega^2-v_s^2k^2]^{\frac{\beta^2}{4\pi}-1}$. If we now assume $v_s$
 is renormalized to zero, we get $ [\omega^2]^{\frac{\beta^2}{4\pi}-1}$

\begin{figure}
\centering
\includegraphics[scale=.4]{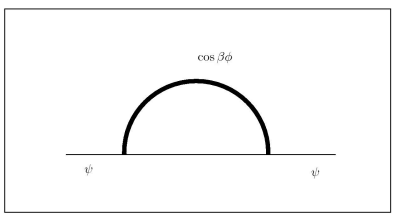}
\caption{1 loop correction to $\psi$ propagator}~\label{1loopBFT}
\end{figure}

Thus a one loop correction (Fig.(\ref{1loopBFT})) to the $\psi $ two point function (inverse propagator) is:
\be
\Sigma (p)\approx \int \frac{d\omega d k}{(2\pi )^2} {(\omega +v_F k) \over (\omega^2-v_F^2k^2) ((p^0-\omega)^2)^{{\beta^2 \over 4\pi}-1}}
\ee
We can do the $k$ integral to get $ln~\omega$ and then do the $\omega$ integral to get
$\Sigma (p^0,p^1)\approx {[ln~p^0 -C - \psi (p^0)]\over (p^0)^{{\beta^2\over 4\pi}-3}}$.
This is a $p^1$ independent non-analyticity that one is seeing in the AdS computation. 
The crucial element that made this happen was the fact that the velocity of the $\Psi$ was  zero.
 This could correspond to some effective non propagating or localized particle.

\section{Conclusion}

We have attempted to understand the behavior of Green's functions of a 1+1 field theory with some background charge
density using the AdS/CFT correspondence. The forms of Green's functions of scalars, fermions and currents have been obtained
at zero as well as non-zero temperatures. The fermion Green's function shows interesting behavior - not typical
of a Landau Fermi liquid and can be termed non-Fermi liquid. In 1+1  dimensions one may expect Luttinger liquid behavior
under some circumstances. At high frequencies this is seen. However at low frequencies the behavior is quite different and seem to have similarities with
Fermi-Luttinger Liquid.
We have also made detailed comparison for $T=0,~~ \omega \rightarrow 0$ with expectations from $AdS_2$ and found 
reasonable agreement. In particular the intriguing log periodicity found in \cite{Faulkner09,Liu09} is also seen. We have also suggested a possible explanation for the $k$-independent non-analyticity at $\omega=0$ in the fermion (or scalar) 
Green's function from the point of view of the boundary theory. 
The non-analyticity can be explained if there are modes with almost zero velocity (i.e. non propagating) that interact with these fermions. These could be impurities for instance.

Finally there is the  obvious question of understanding what experimental setup would correspond to these theories. Also
because the boundary is 1+1 dimensional, more detailed theoretical calculations may be tractable. This would be a good way
to test the AdS/CFT correspondence in more general situations such as the one discussed in this paper. 
 
\noindent
{\bf Acknowledgements}\\
We would like to thank  T. Senthil, T Vojta and especially G. Baskaran for  extremely useful discussions.

\section*{Appendix}

\appendix
\section{The Background Geometry}\label{AppendixMetric}

A solution $2+1$ dimensional Einstein-Maxwell action (\ref{EMAction}) is given  by the following metric(~\cite{Teitelboim99},~\cite{Cadoni08})
\be
ds^2 = -r^2 f(r) dt^2 + \frac{dr^2}{r^2 f(r)}+ r^2 d\theta^2
\ee
where, $f(r)=\frac{1}{l^2}-\frac{8 G M}{r^2}-\frac{8 \pi G Q^2}{r^2} ln(\frac{r}{l})$ and 
$F_{tr}=\frac{Q}{r}$. Also let $r_+$ be the horizon radius determined by the largest
real root of $f(r_+)=0$. The coordinates have following range: $-\infty <t< \infty$ , 
$r_+\le r < \infty $ and $0 \le \theta < 2 \pi$. 
We  scale the coordinates and redefine parameters in the following way 
such that $\theta$ can be replaced by $x$, where $-\infty <x< \infty$,
\be
r = \lambda r' ,\ t = \frac{t'}{\lambda}, \lambda d\theta = {dx\over l}, \ Q = \lambda Q', 
\ M = \lambda^2 (M' - \pi Q'^2 ln~\lambda )
\ee
We can also define $r_+= \lambda r_+'$.
 Then the  metric is given by ,
\be \label{unscmetric}
ds^2 = -r'^2 f(r') dt'^2 + \frac{dr'^2}{r'^2 f(r')}+ {r'^2\over l^2}dx^2
\ee
with $f(r')=\frac{1}{l^2}-\frac{8 G M}{r'^2}-\frac{8 \pi G Q'^2}{r'^2} ln(\frac{r'}{l})$. Note that
$f(r_+')=0$ now, and the vector potential is given by,
\be
A=-Q' ln(\frac{r'}{r_+'}) dt'
\ee
We have chosen a gauge such that potential at horizon is zero. We now introduce a further 
redefinition of coordinates to make all the parameters and coordinates dimensionless\footnote{Note that the parameter $Q$ has dimensions $length^{-{1\over 2}}$}.  This is convenient for calculations.

Thus consider the redefinitions:
\begin{eqnarray}
r' = r'_+ r'' , \quad Q' = \frac{r'_+}{l \sqrt{16 \pi G}} Q'', \nonumber \\
(t',x') = \frac{l^2}{r'_+} (t'',x''), \quad A'_0 = \frac{r'_+}{l \sqrt{16 \pi G}} A''_0
\end{eqnarray}
After which the metric becomes (dropping primes for simplicity),
\be
\frac{ds^2}{l^2}=-r^2 f(r) dt^2 + \frac{dr^2}{r^2 f(r)}+r^2 dx^2
\ee
where,
\be
f(r)=(1-\frac{1}{r^2}-\frac{Q^2}{2 r^2}ln(r))
\ee
and,
\be
A=-Q ln(r) dt
\ee
We now choose a coordinate $z=\frac{1}{r}$ and set $l=1$, we get the metric (\ref{metric})

\section{Spinor Green's Function}\label{AppendixGreF}
The action for bulk spinor field is given by,
\be~\label{Dac}
 S_{spinor} =  \int d^{3} x \sqrt{-g} \, i (\bar \Psi \Gamma^M {\cal D}_M \Psi  - m 
\bar{\Psi} \Psi)
\ee
where 
\be
\bar{\Psi} = \Psi^\dagger \Gamma^{\ud t}, \quad {\cal D}_M  = \partial_M + {1 \over 4} 
\omega_{ab M} \Gamma^{ab} - i q A_M
\ee
and $\omega_{ab M}$ is the spin connection\footnote{We will use $M$ and $a,b$ to denote bulk
space-time and tangent space indices respectively, and $\mu, \nu \cdots$ to denote indices along
the boundary directions, i.e. $M = (z, \mu)$.The $\Gamma^a$ obey 
$\{ \Gamma^a, \Gamma^b\}=2 \eta^{a b}$, where $\eta=diagonal(-1,1,1)$ and 
$\Gamma^M = e^M_{a} \Gamma^a$, where $e^M_{a}$ is the vielbein. Also $(t,z,x)$ denote
space time indices and  $(\ud t,\ud z, \ud x)$ the tangent space indices.}. The non-zero 
components of spin connection are given by,
\begin{eqnarray}
\omega_{{\ud t}\, {\ud z}}&=(\frac{f}{z}-\frac{f^{\prime}}{2})dt \nonumber \\
\omega_{{\ud z}\, {\ud x}}&=\frac{\sqrt{f}}{z} dx
\end{eqnarray}

To analyse Dirac equation following from (~\ref{Dac}) in the background (~\ref{metric}), we use the following basis,

\begin{eqnarray}
&\Gamma^{\ud t}=i \sigma_2 
= \left( \begin{array}{c c}
0 & 1 \\
-1 & 0 \end{array}\right) ,\quad 
\Gamma^{\ud x }= \sigma_1 
= \left( \begin{array}{c c}
0 & 1 \\
1 & 0 \end{array}\right) ,\nonumber \\
&\Gamma^{\ud z}= \sigma_3 
= \left( \begin{array}{c c}
1 & 0 \\
0 & -1 \end{array}\right),\quad
 \Psi=\left( \begin{array}{c} \Psi_+ \\ \Psi_- \end{array} \right)
\end{eqnarray} 
where $\Psi_{\pm}$ are complex functions. Now writing
\be
\Psi_{\pm}=e^{-i \omega t+ i k x} \ \psi_{\pm}(z)
\ee
the Dirac equation becomes,
\begin{eqnarray}
(z f(z) \partial_z - f(z) + \frac{1}{4} z f'(z) \mp m \sqrt{f(z)} ) \psi_{\pm}  \nonumber \\
 = i z (\omega + \mu \ ln(z) \mp k \sqrt{f(z)} ) \psi_{\mp}
\end{eqnarray}
where $\mu=qQ$. 
Let us consider the following transformation
\be
\tilde{\psi}_{\pm}=e^{-\int_a^z \frac{dt}{t f(t)}\left[ f(t)-\frac{1}{4} t f'(t) \right]} 
\psi_{\pm}
\ee
where $a$ is some constant. The equations for $\tilde{\psi}_{\pm}$ becomes,
\be~\label{psifo}
\left[ z f(z) \partial_z  \mp m \sqrt{f(z)} \right] \tilde{\psi}_{\pm} 
 = i z (\omega+ \mu \ ln(z) \mp k \sqrt{f(z)} ) \tilde{\psi}_{\mp}
\ee
Using (~\ref{psifo}) we can get the equation for $G(z)$ given by  (\ref{Fgreenseq}). We will also drop the tilde for simplicity.
Note that in (~\ref{psifo}),

In order get the retarded correlation function for the dual fermionic operator
${\cal O}$, we need to impose ingoing boundary condition for $\psi$ 
at the horizon. 

We substitute $\psi_+(z)=e^S$ in (\ref{psifo}) to obtain the leading behavior as $z\rightarrow 1$,
\be  \label{B1}
\psi_- ={z f S'-m \sqrt{f} \over iz(\omega+\mu~ln(z)-k \sqrt{f})}e^S
\ee
Assume that the dominant term is the one with the 
$z$-derivative acting on $e^S$. This gives
\be
\partial _z \psi _- = S'{z f S'- m \sqrt{f}\over iz(\omega +\mu ~ln~z - k\sqrt{f})}e^S
\ee
Plug this into other equation of (\ref{psifo}) to get an equation for $S$, which is given by:
\be
S'(z)=\pm {\sqrt{m^2 + z^2 (k^2-({\omega +\mu~ln~z\over \sqrt{f}})^2)}\over z \sqrt{f}}
\ee
and,
\be \label{gz}
G(z) = \frac{z f S' -m \sqrt{f}}{z (\omega+ \mu ln(z)-k \sqrt{f})}
\ee
This can be used to derive the asymptotic behavior (near $z=1$) of the Green's functions. \\
{\bf Case 1} $\mathbf{Q^2 \ne 4}$:  $f(z) \approx 4 \pi T (1-z)$, and $\psi_+ = e^{S(z)} \approx e^{\mp i \frac{\omega}{4 \pi T} ln(1-z)} $. The negetive sign in the exponent gives
in-going wave. Putting back $S(z)$ with this choice of solution in equation (\ref{gz}), gives $G(z=1)=i$.
\\
{\bf Case 2} $\mathbf{Q^2 = 4}$: In the extremal case where $f(z) \approx 2 (1-z)^2$ , we get
\be
S'(z)=\pm {\sqrt{m^2 + k^2-{1 \over 2}({\omega\over (1-z)}-\mu)^2}\over  \sqrt{2} (1-z)} 
\label{B4}
\ee 
for $\omega \neq 0$, we can neglect the other terms (keeping next to leading order term),
\be
\psi_+ =(1-z)^{\pm {i\mu\over 2}}e^{\pm {i\omega \over 2 (1-z)}}
\ee
The positive sign in the exponent is the ingoing wave. Also near $z=1$ using (\ref{gz}) we conclude that near $z=1$, for ingoing waves,
\be
G= i {\psi _- \over \psi_+}={\sqrt{m^2+k^2-{1 \over 2}({\omega
\over (1-z)}-\mu)^2}-m \over {1 \over \sqrt 2} ({\omega \over (1-z)}-\mu -k \sqrt 2)}
\ee
This gives the boundary condition for $\omega \ne 0$
\be G(z=1)=i\ee

In the case where $\omega =0$ we get,
\be
G(z=1)= {m- \sqrt {m^2 + k^2-{\mu^2\over 2} -i \epsilon}\over  (k+{\mu\over \sqrt2})}
\ee
where $m^2 \to m^2 - i \epsilon$ to choose the correct sign in square root.

\section{Vector Green's Function}\label{AppendixGreV}
The perturbations in $A$ and $g_{tx}$, which we call $a(z,t)$ and $\delta (z,t)$ respectively obeys (we are assuming a time dependence of $e^{-i\omega t}$ for both $a$ and $\delta$) following set of equations:
\begin{eqnarray}~\label{adelta}
& f \partial_z (z f \partial_z a)+z \omega^2 a + Q f \partial_z (z^2 \delta) =0\nonumber \\
& z Q a + \partial_z(z^2 \delta)=0
\end{eqnarray}
Where the $f(z)$ is given by (\ref{metric}).
We can combine the two equations (\ref{adelta}) to get an equation for $a(r,t)$ - which is:
\be
f^2 a'' + f ( {f \over z} +f')a' + (\omega^2 - 2Q^2 f)a =0
\ee

It is also convenient to factorize the oscillatory behavior at the horizon where $f(z)=0$. Thus we can define $a(z)= (f(z)^{i\omega \over c})\rho(z)$.  The differential equation for $\rho(z)$ is:
\beqa
&& f^2 \rho'' + [ \frac{f^2}{z}+(1+ 2 i {\omega \over c}) f f' ] \rho' 
+  [ {i \omega \over c} f f'' \nonumber  \\
&& + {i \omega \over c z} f f' - {\omega^2 \over c^2} f'^2 + \omega^2-Q^2 f ] \rho =0
\eeqa 
$c$ gets fixed by requiring that $\rho(1)$ be non-zero. One finds $c^2= (f'(1))^2$. The sign is fixed by requiring that the solution for $a$ be ingoing. This gives $c=-f'(1)$. 
\subsection{Comments on Vector Green's function}
We consider the case of the  scalar field with degenerate eigenvalues. Thus if $f_1(z) \approx z^\lambda$ is a solution then the Frobenius prescription is to consider ${d f_1\over d~\lambda}$ as the second solution. 
So we have a situation where $\phi (z) \approx  a^+ z^\lambda + a^-  z^\lambda ln~z +...$ is the form of the solution where the dots indicate terms that are determined algebraically in terms of $a^{\pm}$.
 We have to decide which solution corresponds to the source and which one to the expectation value. The original idea was that one solution (non-normalizable) solution corresponds to the source and the other (normalizable) one gives the expectation value.
 This idea has been extended and it has also been pointed out that in some situations when both are normalizable one can have a duality between the two possibilities and that both are valid. In our case the solution $z^\lambda ln~z$ is more divergent and
 is best taken as the source. In that case the Green's function is $a^+\over a^-$. One can consider the case where we have $\lambda ^\pm$ wit $\lambda ^+ \approx \lambda ^-$. 
Then take the two independent solutions to be $a^+ z^{\lambda ^+}$ and $a^-{z^{\lambda ^+}-z^{\lambda ^-}\over \lambda ^+ -\lambda ^-}$ - anticipating that we are going to set $\lambda^+=\lambda ^-$. 
If one then applies the prescription of ~\cite{Son02} with $\phi $ normalized to one at the boundary one gets, on setting $\lambda^+=\lambda ^-=\lambda$,
\[
G={\partial_ z \phi (\epsilon)\over \phi (\epsilon)}
\]
\be
={a^+ \lambda + a^- - a^- \lambda ln~\epsilon\over a^+ + a^- ln~\epsilon}
\ee
When $\epsilon \rightarrow 0$ we get a leading piece $\lambda$, which is uninteresting because it is analytic in momenta, and the sub leading non analytic 
piece $\lambda a^+\over a^-$. In case $\lambda =0$ we get
\be
{ a^- \over a^+ + a^- ln~\epsilon}={ a^- \over ({a^+\over  a^- ln~\epsilon} +1) a^- ln~\epsilon}
\ee
So the leading non-analytic piece is again ${1\over ln^2~\epsilon}{a^+\over a^-}$. Thus we can conclude that in general for degenerate cases $G={a^+\over a^-}$.

\subsection{Scalar Green's function}\label{AppendixGreS}

Consider the equation of motion for the scalar field (\ref{scalareqn}).
To minimize errors in numerical computations it will be helpful to factor out the oscillations near the
horizon be writing $\phi_k(z)=f(z)^{\nu}S(z)$. By expanding the differential equation about
the horizon value of $z$ one can fix $\nu$ to be $-i\omega/4\pi T$ where, $T$ is the temperature
of the black hole obtained in section 2. The minus sign is chosen for the ingoing solution at the horizon. 
The resulting differential equation in $S(z)$ is,
\beqa
&& S'' Af^2+S'(2\nu A ff'+Bf^2)+ \nonumber \\
&& S((\nu(\nu-1)Af'^{2}+\nu Aff''+\nu B f f' +Vf^{2})=0
\eeqa 

The solution to the equation (\ref{scalareqn}) near the boundary, $z\rightarrow 0$, for non-zero mass of the scalar field 
behaves as, (we have dropped the log term coming from the potential term, $V(z)$ in (\ref{scalareqn}) assuming a finite mass for the scalar field\footnote{In the following numerical computations we have set $m^2=1/4$.}) 

\beqa
\phi(z)=a^-(k,\omega)z^{d-\Delta}+a^+(k,\omega)z^{\Delta}
\eeqa

Here $d=2$ is the boundary space-time dimension and $\Delta=d/2+\sqrt{(d/2)^2+m^2}$. As per \cite{Kleb-Witten} the boundary
Green's function is given by, $a^+/a^-$. The solution for $\phi(z)$ is obtained numerically and 
the boundary Green's function can be written as (\ref{gfscalar}).

For the extremal case, ($Q=2$) a slight modification of the above procedure is needed due to the presence of degenerate horizons.
We write $\phi_k(z)=e^{\nu_1/(1-z)}(1-z)^{\nu_2}S(z)$, where $\nu_1=i\omega/2$ and $\nu_2=i(q-\omega/6)$. The
signs are fixed so that the infalling solution is chosen at the horizon. Again this
is done to factor out the oscillations near the horizon. Like the finite temperature case the resulting differential equation  
(\ref{diffzerot}) can now be solved numerically. The boundary Green's function is given by (\ref{gfscalar}).
\beqa\label{diffzerot}
S^{''}A (z-1)^4 &+&
S^{'}[ (z-1)^2\{2 A (\nu_1  + \nu_2 (z-1))+B (z-1)^2\}]+\nonumber\\
S [ A \{\nu_1^2 &+& 2 \nu_1 (\nu_2-1) (z-1) + (\nu_2-1) \nu_2 (z-1)^2\} \\ &+&
B\{(z-1)^2(\nu_1 + \nu_2 (z-1))\} + V(z-1)^4 ]\nonumber
=0
\eeqa

\section{$AdS_2$ limit}\label{AppendixAdS2}
As is described in ref.(\cite{Faulkner09},\cite{Faulkner10}), at $T=0$ the low frequency limit divides the full bulk space-time into two distinct regions. Near the  horizon called inner region we have an $AdS_2$ factor where as outer
region is described by $AdS_3$. For inner region, we define  new coordinates $(\tau, \zeta)$ such that, 
\be \label{scale}
1-z= \frac {\omega} {2 \zeta}~~~~:~~~~~\tau= \frac 1 {\omega} t
\ee
such that $\omega, (1-z) \to 0$, both $\zeta$ and $\tau$ are finite. In this limit the background metric (\ref{metric}) in new co-ordinates is $Ads_2 \times \mathbb{R}$,
\be \label{nhmetric}
ds^2 = \frac 1{2 \zeta^2} (- d\tau^2 + d\zeta^2 )+ dx^2~~~~;~~~~
A_{\tau}=-\frac{1}{\zeta}
\ee
where, radius of $AdS_2$ subspace $R_2 = 1/\sqrt{2}$ in the unit of  $AdS_3$ radius which we have set to unity. A  bulk field $\Phi$ in $AdS_3$ maps to a tower of
fields $\Phi_k$, where $k$ is the momentum along $\mathbb{R}$. Similarly the boundary operator $\mathcal{O}(k)$ maps to an IR operator $\mathcal{O}^{IR}_k$ in IR CFT. The Green's function ($\mathcal{G}_k(\omega)$) 
for IR operator  dual to $\Phi_k$ can be obtained analytically by solving bulk equation of motion of $\Phi_k$ in $AdS_2$ with constant electric field (\ref{nhmetric}). Small $\omega$ limit of the full Green's function can be obtained by matching
the inner and outer solution. The perturbative expansion for the retarded Green's function of full theory can be written as for a generic value of $k$,
\be \label{pertgreens}
G_R(\omega,k)= (a_0(k)+\omega a_1(k) +{\cal O}(\omega^2)) + {\cal G}_k(\omega)
\left(b_0(k)+\omega b_1(k) +{\cal O}(\omega^2)\right),
\ee
where, $a$ and $b$ are real function of momentum $k$ and $\mu$ obtained by matching solutions of inner and outer region. So the imaginary part of $G_R$ or the spectral function is controlled by $\mathcal{G}_k$.
We study the three cases spinor,vector and charged scalar in turn.

\subsection{Fermion}
Near horizon, zero frequency limit of  (\ref{psifo}) in new co-ordinates (\ref{scale}),
\be
\left[ 2 \zeta \partial_{\zeta}  \mp m \sqrt{2} \right] \tilde{\psi}_{\pm}
 = i (2 \zeta - \mu \mp k \sqrt{2} ) \tilde{\psi}_{\mp}
 \ee
where $\mu=2q$.
These equations are basically the equations for a fermion in $AdS_2$ with an additional "mass" term in the 2-D action of the form $ik\bar \psi\Gamma \psi$ where $\Gamma =\sigma _1$.
We can analytically calculate $Ads_2$ dual CFT correlation function from the above equation \cite{Faulkner09}.  
The asymptotic behavior of the above set of equations can be obtained
by taking ($\zeta \rightarrow 0$) giving
\be
\left[ 2 \zeta \partial_{\zeta}  \mp m \sqrt{2} \right] \tilde{\psi}_{\pm}
 = i (- \mu \mp k \sqrt{2} ) \tilde{\psi}_{\mp}
  \ee
The solutions are
\begin{eqnarray}
\psi_+& =& C \zeta^{-\nu_k} + \tilde{C} \zeta ^{\nu_k}\\
\psi_-&=&\frac i {2}\left[- 
\frac {C (2 \sqrt{2} m + 4 \nu_k)
\zeta^{-\nu_k}}{(\mu + \sqrt{2}k)} + \frac 
{\tilde{C}(- 2 \sqrt{2} m + 4 \nu_k)  
\zeta^{\nu_k}}{(\mu + \sqrt{2}k)}\right],
\end{eqnarray}
where, 
\be
\nu_k = \frac 1 2 \sqrt{2(m^2 + k^2) - \mu^2},
\ee 
$C_1$ and $ C_2$ are the integration constants.The scaling dimension of the operator $\mathcal{O}_k$ is $\Delta=1/2 + \nu_k$. So, the leading order $\omega $ dependence (going back to $z$ coordinates) 
is $\omega^{\pm \nu_k}$. Using  which we can get  the expression for IR Green's function 
\be \label{FGreensAds2}
{\cal G}_k(\omega) =  \langle{\cal O}_{k'}(\omega) {\cal O}_{k}(\omega)\rangle \propto \delta(k-k') {\cal H}(\mu,k) \omega^{2 \nu_k},
\ee

It is possible for fermionic case that for specific values of $k$ (determined numerically, denoted as $k_F$) the perturbative expansion for $G_R$ takes a form different from (\ref{pertgreens}),
\be \label{pertgreenskf}
G_R(k,\omega)=\frac{h_1}{k-k_F-\frac{\omega}{v_F}-\Sigma[\omega,k]}
\ee 
where $\Sigma[\omega,k]=h {\cal G}_{k_F}(\omega) $. Again  the spectral function is controlled by ${\cal G}_{k_F}(\omega)$. If such $k_F$ exists, then there is a pole at $k=k_F$ and $\omega=0$ and can be identified as quasiparticle pole.
But generically for any $k$, $Im(G_R)$ goes to zero if $\nu_k$ is real, or oscillatory if $\nu_k$ is purely imaginary in the small $\omega$ limit.

\subsection{Vector}
As we have discussed extensively for the fermionic case, for this 
case also we do the same co-ordinate transformation for 
the near horizon limit of the extremal black hole. So, in the 
near horizon $AdS_2$ the equation of motion for the gauge field,

\be
f(z) \partial_z (z f(z) a') + z (\omega^2 - 4 z^2 f(z)) a =0 
\ee
turns out to be 

\be
\zeta^2 \frac{d^2 a}{d \zeta^2} + (\zeta^2 -2)a =0,
\ee
This equation is same as the equation for a scalar with $m^2=2$ and no charge. 
This can be analytically solved. The solution with ingoing boundary condition
at the horizon becomes,
\beqa
a = C \sqrt{\frac 2 {\pi}}\left [ i \left(- \cos(\zeta)+ 
\frac {\sin(\zeta)} \zeta \right) + \left(\sin(\zeta) - 
\frac {\cos(\zeta)} \zeta \right)\right],
\eeqa
where $C$ is the integration constant.
Solution near the boundary of $AdS_2$ $\zeta\rightarrow 0 $ looks like
\be
a= -\frac 1 {\zeta} - \frac i 3 \zeta^2
\ee
So, ofter going back to the original co-ordinate the lading leading order
$\omega$  behavior of the IR Green's function would be ${\cal G}_R (\omega) = \frac{i}{3}  \omega^3 $.

\subsection{Charged Scalar}
The equation of motion of charged scalar field (\ref{scalareqn}) in $AdS_2$ limit becomes,
\be
\frac {d^2 \phi}{d \zeta^2} + \left[\left(1- \frac {\mu}{2 \zeta}\right)^2
- \frac 1{2 \zeta^2}(m^2 + k^2)\right]\phi=0,
\ee
where $\mu=2q$.
It is important to note that the above equation is same as a  
charged scalar field equation of mass $m^2_k= m^2 + k^2$ in 
 a background $AdS_2$ space (\ref{nhmetric}). 

The asymptotic behavior of this equation may be 
obtained by taking $\zeta \rightarrow 0$ and one finds
\be
\phi (r) \approx  A \zeta^{\lambda_+} +B \zeta^{\lambda_-}
\ee
where $\lambda$ are eigenvalues of
\beqa
\lambda (\lambda -1) +\frac 1 4 [\mu^2 - 2(m^2 + k^2)]=0  \nonumber \\
\implies 
\lambda_\pm = {1\over 2}\pm \frac 1 2 \sqrt{1 + 2(m^2+k^2)-\mu^2}
\eeqa
If we now substitute $\zeta= \frac {\omega}{2(1-z)}$,
we get  
\be
\phi (z) \approx A \omega^{\lambda _+}(1-z)^{-\lambda _+} + 
B \omega^{\lambda_-}(1-z)^{-\lambda_-}
\ee
From this we can conclude that the Green's functions in the $AdS_2$ (ratio of A/B) goes as $\omega ^{\lambda_+-\lambda _-}= \omega ^{2\nu }$ where $\nu = \sqrt{{1\over 4}+{(m^2+k^2 \over 2}-{\mu^2\over 4})}$ .

\end{document}